\documentclass{aa}  

\usepackage{graphicx}
\usepackage{multirow}
\usepackage{arydshln}
\usepackage{adjustbox}
\usepackage{amsmath}
\usepackage{txfonts}
\usepackage{hyperref}
\usepackage{float}
\hypersetup{
    linkcolor=blue,
    citecolor=blue,
    colorlinks=true,
    filecolor=blue,      
    urlcolor=blue}

\begin{document} 

   \title{Correcting the hydrostatic mass for non-thermal gas motions:\\ a comparison of two approaches}

   \titlerunning{Virgo replica mass estimate correction}

    \author{Th\'eo Lebeau\inst{1,2,3} \thanks{Corresponding author: \href{mailto:tlebeau@astro.rug.nl}{tlebeau@astro.rug.nl}}, Nabila Aghanim\inst{3}, Jenny G. Sorce\inst{4,3}, Stefano Ettori\inst{5,6}}

   \authorrunning{Lebeau et al.}

   \institute{Kapteyn Astronomical Institute, University of Groningen, Groningen, The Netherlands
        \and 
            Astrophysics Research Center of the Open University (ARCO), The Open University of Israel, Israel
        \and 
            Université Paris-Saclay, CNRS, Institut d’Astrophysique Spatiale, 91405 Orsay, France
        \and
            Université de Lille, CNRS, Centrale Lille, UMR 9189 CRIStAL, F-59000 Lille, France
        \and
            INAF, Osservatorio di Astrofisica e Scienza dello Spazio, via Piero Gobetti 93/3, 40129 Bologna, Italy
        \and
            INFN, Sezione di Bologna, viale Berti Pichat 6/2, 40127 Bologna, Italy}

   \date{Received XXX / Accepted XXX}

\abstract
{An accurate estimation of the mass of galaxy clusters is key to precisely and unbiasedly constraining cosmological parameters through their number count. The hydrostatic mass, estimated from the properties of the intracluster medium (ICM) assuming hydrostatic equilibrium, sphericity, and thermal-only pressure, is known to be biased by 10 to 20\%, most likely due to non-thermal pressure support from gas motions. Two corrections have been proposed: i) replacing the thermal pressure by the total pressure $P_\mathrm{tot}=P_\mathrm{th}+P_\mathrm{nth}$, or ii) adding effective mass terms derived from the gas momentum equation. We compare these approaches using a numerical replica of the Virgo cluster as a case study, estimating corrected masses from 3D radial profiles in different cluster regions and from projected sightline velocities mimicking XRISM observations. We find that the two methods do not yield the same results in 3D: the non-thermal pressure correction increases the mass by a growing amount with radius (from a few per cent in the core to $\sim$40\% at the virial radius), whereas the effective mass terms provide a correction that varies less with radius. When estimated from projections, the two methods agree to within a few per cent for a given sightline, but the non-thermal pressure fraction is underestimated by about a factor of 2 compared to the 3D case. Furthermore, projection effects can change the inferred non-thermal pressure fraction by up to a factor of 2, particularly when the sightline is aligned with cosmic filaments.}


\maketitle

\section{Introduction}

The population of galaxy clusters and its evolution in mass and redshift directly depend on the $\Omega_m$ and $\sigma_8$ cosmological parameters, making them powerful cosmological probes \citep[see e.g.][for a review on cosmological parameters]{lahav2022cosmologicalparameters2021}. These objects are indeed located at the nodes of the cosmic web, corresponding to density peaks in the Universe's matter distribution. They originate from primordial overdensities that have continuously accreted matter since the earliest stages of cosmic evolution \citep[see e.g.][]{liddle2015introduction,peebles2020large}. The key step in using galaxy clusters as cosmological probes through their number counts is to calibrate their mass in an unbiased way \citep[see, e.g.,][for reviews]{allen2011cosmological,pratt2019galaxy}. Several methods can be used to compute the cluster mass, among which are scaling relations based on the self-similar model of \citet{kaiser1986evolution} (see e.g.   \citeauthor{2013SSRv..177..247Giodini} \citeyear{2013SSRv..177..247Giodini} for a review and
\citeauthor{2024A&A...690A.238Aymerich}
\citeyear{2024A&A...690A.238Aymerich},\citeyear{2025arXiv250902068Aymerich} and references therein for recent
works), the spatial and velocity distribution of cluster galaxies (see for example {\tt Caustic} in \citeauthor{diaferio1997infall} \citeyear{diaferio1997infall} or {\tt MAMPOSSt} in \citeauthor{mamon2013mamposst} \citeyear{mamon2013mamposst}), or weak lensing (see \citeauthor{hoekstra2013masses} \citeyear{hoekstra2013masses} for a review and \citeauthor{okabe2016locuss} \citeyear{okabe2016locuss}, \citeauthor{applegate2014weighing} \citeyear{applegate2014weighing}, \citeauthor{umetsu2016clash} \citeyear{umetsu2016clash}, and \citeauthor{herbonnet2020cccp} \citeyear{herbonnet2020cccp} respectively for the LoCuSS, WtG, CLASH and CCCP catalogues).

Another way is to estimate the total mass of galaxy clusters from the properties of the intracluster medium (ICM) by assuming hydrostatic equilibrium, that is, the balance between the thermal ICM pressure and the cluster gravitational potential, assuming spherical symmetry \citep[see, e.g.,][for a review]{ettori2013mass}. This hydrostatic mass has been estimated from observations in the X-rays \citep[e.g.][]{2005A&A...435....1Pointecouteau,vikhlinin2006chandra,2009ApJ...692.1060Vikhlinin,2009A&A...498..361Pratt,2019A&A...621A..39Ettori} but has been shown in simulations to be biased by 10--20\% due to non-thermal pressure support \citep[see e.g.][]{rasia2006systematics}, a result that has been confirmed and further quantified observationally \citep[e.g.][and references therein]{eckert2019non}. This hydrostatic mass bias is nowadays calibrated from simulations and accounted for in the cosmological analysis, though with values distributed around $(1-b) \simeq 0.8$ among the different studies \citep[see comparisons in e.g.  ][]{biffi2016nature,gianfagna2021exploring,2024A&A...682A.157Lebeau}. Moreover, the value proposed in \citet{salvati2018constraints} to reduce the tension between constraints on $\Omega_m$ and $\sigma_8$ from Cosmic Microwave Background (CMB) and cluster number counts from thermal Sunyaev--Zel'dovich \citep[tSZ,][]{sunyaev1972observations} observations is $(1-b) = 0.62 \pm 0.07$, which is in disagreement with most of the values calibrated from simulations and weak-lensing. There is thus a need for a better understanding and correction of the physical sources of deviation from hydrostatic equilibrium in the ICM.

As a matter of fact, various non-thermal processes \citep[see e.g.][for a review]{2012IJMPS..12..280Ferrari} could contribute to push the ICM out of hydrostatic equilibrium, such as magnetic fields (see e.g. \citeauthor{2005JCAP...01..009DolagUHECRs} \citeyear{2005JCAP...01..009DolagUHECRs}, \citeyear{dolag2008simulation}) and cosmic rays \citep[see e.g.][for a review]{2019SSRv..215...16VanWeeren}. However, the largest contribution to non-thermal pressure is thought to arise from chaotic motion of the gas generated by processes such as accretion, merging, or stellar and Active Galactic Nuclei (AGN) feedback, which drive shocks and hydrodynamic instabilities and ultimately lead to turbulence \citep[see e.g.][for a review]{2019SSRv..215...24Simionescu}. Analytical models for the resulting non-thermal pressure have also been developed \citep[e.g.]{2014MNRAS.442..521Shi}, including a proposed correction of the hydrostatic mass bias \citep[e.g.]{2016MNRAS.455.2936Shi}. There is indeed growing evidence for turbulence in and around galaxy clusters, both from simulations \citep[e.g.][and references therein]{rasia2004dynamical,nelson2014hydrodynamic,gaspari2013constraining,2020MNRAS.495..864Angelinelli,valles2021troubled,2025A&A...693A.263Groth,2025A&A...704A..14Lebeau,2026A&A...707A.336Lebeau} and observations, either through high-resolution spectroscopy with XRISM \citep[including e.g.][]{2025ApJ...982L...5XRISM,2025ApJ...993L..11XRISM_comp,2025ApJ...985L..20XRISM-Coma,2025PASJ...77S.242XRISMA2029}, or using fluctuations of thermodynamical properties of the ICM \citep[see e.g.][]{schuecker2004probing,2023A&A...673A..91Dupourque,dupourque2024chex,romero2023inf,romero2024surface,2024A&A...682A..45Lovisari,2025A&A...694A.182Adam}. Moreover, an important predicting work is being carried out in prevision of the future with NewAthena sattelite \citep[e.g.][]{2019A&A...629A.143Clerc,2024A&A...686A..41Beaumont,2025A&A...702A.215Molin}.

In that context, correction methods to account for gas motion in the mass estimation of clusters have been proposed. On the one hand, one can directly correct the hydrostatic mass by adding the non-thermal pressure contribution to the total pressure support \citep[as in e.g.][]{nelson2014hydrodynamic,pearce2020hydrostatic,gianfagna2021exploring,2022A&A...657L...1Ettori}. On the other hand, one can account for gas motion by adding additional effective mass terms derived from the Euler gas momentum equation \citep[as in e.g.][]{2009ApJ...705.1129Lau,2013ApJ...777..151Lau,2013ApJ...767...79Suto}. In this work, we compare these two approaches using a numerical replica of the Virgo cluster \citep{sorce2021hydrodynamical} as a case study. We first provide a concise description of the simulation in Sect.~\ref{sec:2 Virgo} and present the hydrostatic mass and its correction accounting for the velocity field in Sect.~\ref{sec 3: mass theory}. We then estimate these masses and their bias with respect to the total mass using 3D radial profiles in Sect.~\ref{sec4: 3d mass}. Using projected quantities, we explore possible estimates of the hydrostatic mass and its correction in the case of XRISM exposures in Sect.~\ref{sec:2d masses}. We finally discuss our results in Sect.~\ref{discussion} and conclude in Sect.~\ref{conclusion}.

\section{The Virgo cluster simulated replica}
\label{sec:2 Virgo}

The present study is based on a constrained hydrodynamical zoom-in simulation of the Virgo cluster embedded in its large-scale environment; full details of its construction and validation are provided in \cite{sorce2021hydrodynamical}, which was extensively used in \cite{2024A&A...689A..19Lebeau,2024A&A...682A.157Lebeau,2025A&A...704A..14Lebeau,2026A&A...707A.336Lebeau,sorce2026ii}. We here summarise the main characteristics of this numerical replica of the Virgo cluster. The simulation is part of the CLONE (Constrained LOcal and Nesting Environment) project, in which the initial conditions of the local Universe are reconstructed from observed galaxy positions and peculiar velocities taken from the Cosmicflows-2 catalog \citep{tully2013cosmicflows} via a reverse Zel'dovich approximation following the method described in \citeauthor{sorce2016cosmicflows} \citeyear{sorce2016cosmicflows}. Among the 200 dark matter replicas of the Virgo cluster produced by \cite{sorce2019virgo} within this constrained environment, the one that best reproduces the statistical properties and merging history of the full ensemble was selected as the basis for the hydrodynamical run \citep{sorce2021hydrodynamical}. The resulting simulation reproduces the observed local environment of Virgo remarkably well, including the foreground and background cosmic filaments along our line of sight and the group of galaxies currently infalling onto the cluster.

The code employed for the simulation is the adaptive mesh refinement (AMR) solver {\tt RAMSES} \citep{teyssier2002cosmological}, run within a 738~Mpc ($500~\mathrm{Mpc\, h^{-1}}$) local Universe box. The zoom-in region centred on Virgo is a sphere of 30~Mpc diameter resolved with an effective number of $8192^3$ dark matter particles, each of mass $m_\mathrm{DM} = 3 \times 10^7~\mathrm{M_\odot}$, and an AMR gas grid reaching a finest cell size of 0.35~kpc in the cluster core. The adopted cosmological parameters follow \cite{planckcosmoparam2014}: $H_0 = 67.77~\mathrm{km\,s^{-1}\,Mpc^{-1}}$, $\Omega_m = 0.307$, $\Omega_\Lambda = 0.693$, $\Omega_b = 0.048$, $\sigma_8 = 0.829$, and $n_s = 0.961$.

The baryonic physics is modeled through sub-grid prescriptions for radiative gas cooling and heating, star formation, and kinetic feedback from type~II supernovae and active galactic nuclei (AGN), following the Horizon-AGN implementation of \cite{dubois2014dancing,dubois2016horizon}. The AGN jet direction is set by the spin of the central black hole \citep{dubois2021introducing}, which produces a more physically motivated feedback geometry than a fixed
axis. Both the dark matter-only and the hydrodynamical runs are in good agreement with observations of the Virgo cluster and its local environment \citep{sorce2021hydrodynamical,2024A&A...682A.157Lebeau}.

\section{Hydrostatic mass and corrections}
\label{sec 3: mass theory}

In this section, we define the hydrostatic mass and the two corrections compared in this work. First of all, the ground-truth total mass, $M_\mathrm{tot}$, used as a reference throughout this work is the sum of the DM particles, $M_{\rm DM}$, and the gas cells, $M_{\rm gas}$, as in \citet{2024A&A...682A.157Lebeau}; the stars-particles represent less than a per cent of the total mass, and are thus neglected. We derive the well-known hydrostatic mass equation starting from the Euler hydrodynamic equation

\begin{equation}
    \frac{\partial \mathbf{v}}{\partial t} + (\mathbf{v} \cdot \nabla)\mathbf{v} = -\frac{1}{\rho_{\rm gas}}\nabla P - \nabla \phi \,
\end{equation}

\noindent with $\mathbf{v}$ the velocity, $\rho_{\rm gas}$ the gas density, $P$ the pressure and $\phi$ the gravitational potential. By assuming a static fluid, or at least that the gas motion has a minor contribution to the ICM dynamics, i.e. $\mathbf{v}=\mathbf{0}$, the hydrostatic equation naturally emerges as

\begin{equation}
    \nabla \phi = -\frac{1}{\rho_{\rm gas}}\nabla P \, . 
\end{equation}

\noindent In the spherical symmetry case, that is $\nabla \phi = \frac{GM}{r^2}$ with $G$ the gravitational constant, $M$ the mass, and $r$ the radius, and $\nabla P=\frac{dP}{dr}$ the radial pressure gradient, the hydrostatic mass thus naturally writes

\begin{equation}
   M_{\rm HE}(<r)=-\frac{r^2}{G\rho_{\rm gas}}\frac{dP}{dr}\,.
\end{equation}

\noindent In addition, we assume the ICM is thermalized; in other words, the thermal pressure, $P_\mathrm{th}$, results from the equilibrium between gravitational heating, radiative cooling, and feedback processes. Consequently, the hydrostatic mass term is labeled as thermal mass, $M_{\rm th}$ \citep[as in other works such as][]{2009ApJ...705.1129Lau}. Moreover, using the perfect gas equation of state, that is $ P_{\rm th}=P_{\rm gas}=(\rho_\mathrm{gas} k_B T)/(\mu m_u)=n_\mathrm{gas} k_B T$ with $\rho_{\rm gas}=n_{\rm gas} \mu m_u$, where $n_{\rm gas}$ is the gas number density, $\mu$ is the mean molecular weight, and $m_u$ is the atomic mass constant\footnote{taken as $\mu=0.6$ and $m_u$ = $1.661\times 10^{-27}~\mathrm{kg}\sim m_p$ in this work.}, $k_{\rm B}$ the Boltzmann constant and $T$ the temperature, we have the usual hydrostatic mass equation

\begin{align}
\begin{split}
   M_\mathrm{HE}(<r)=M_\mathrm{th}(<r) &= -\frac{r P_\mathrm{gas}}{G n_\mathrm{gas} \mu m_u} \frac{d\ln(P_\mathrm{gas})}{d\ln(r)} \\
   &= -\frac{r k_\mathrm{B}T}{G \mu m_u} \left[ \frac{d\ln(n_\mathrm{gas})}{d\ln(r)} + \frac{d\ln(T)}{d\ln(r)}\right] .
\end{split}
\label{equ:mhe_ngas}
\end{align}

If the thermal pressure is taken as the electron pressure, i.e. $P_{\rm th}=P_\mathrm{e} = (\mu/\mu_e) P_{\rm gas}$ and $\rho_{\rm gas}=n_{e} \mu_e m_u$ with $\mu_e$ the mean molecular weight per electron\footnote{taken in this work as $\mu_e=2/(1+X_H)\approx1.1363$ for the fully ionised ICM, with $X_H=0.76$.}, which is the convention adopted in this work, we thus have

\begin{align}
\begin{split}
   M_{\mathrm{HE}}(<r)=M_{\mathrm{th}}(<r) &= -\frac{r P_{\mathrm{e}}}{G n_\mathrm{e} \mu m_u} \frac{d\ln(P_{\mathrm{e}})}{d\ln(r)} \\
   &= -\frac{r k_\mathrm{B}T}{G \mu m_u} \left[ \frac{d\ln(n_\mathrm{e})}{d\ln(r)} + \frac{d\ln(T)}{d\ln(r)}\right]  
   \label{equ:mhe_ne}
\end{split}
\end{align}
\subsection{The non-thermal pressure correction}

In the usual formulation of the hydrostatic mass described just above, only the thermal pressure is accounted for. However, non-thermal pressure in the ICM can arise from cosmic rays, magnetic fields, and non-thermal gas motion. The latter is considered the dominant process and is often assumed to arise mostly from turbulence. In this work, we only consider non-thermal gas motion and thus define the non-thermal pressure as $P_\text{nth}=(\rho\sigma_{\rm 3D}^2)/3$ with $\sigma_{\rm 3D}$ the 3D velocity dispersion, by analogy with the thermal pressure defined with the thermal velocity as $P_\text{th}=(\rho v_{th}^2)/3$. The hydrostatic mass equation can then be corrected by taking the total pressure $P_\mathrm{tot}=P_\mathrm{th}+P_\mathrm{nth}$. Using the non-thermal pressure fraction $\alpha=P_\text{nth}/P_\text{tot}$ as in, e.g., \citet{nelson2014hydrodynamic,2014MNRAS.442..521Shi,shi2016locations,pearce2020hydrostatic,gianfagna2021exploring,2022A&A...657L...1Ettori}, we thus have 

\begin{equation}
\begin{split}
    M_{\mathrm{\alpha}}(<r) &= -\frac{r P_{\mathrm{tot}}}{G n_e \mu m_u} \frac{d\ln(P_{\mathrm{tot}})}{d\ln(r)} \\
   & = -\frac{1}{1-\alpha}\frac{r P_{\mathrm{th}}}{G n_e \mu m_u} \frac{d\ln\left(P_{\mathrm{th}}/(1-\alpha)\right)}{d\ln(r)} \\
   & = -\frac{1}{1-\alpha}\frac{r P_{\mathrm{th}}}{G n_e \mu m_u} \left(\frac{d\ln(P_{\mathrm{th}})}{d\ln(r)}+\frac{\alpha}{1-\alpha}\frac{d\ln(\alpha)}{d\ln(r)}\right) \\
   & = \frac{1}{1-\alpha} \left(M_{\mathrm{th}}(<r)-\frac{\alpha}{1-\alpha}\frac{r P_{\mathrm{th}}}{G n_e \mu m_u}\frac{d\ln(\alpha)}{d\ln(r)}\right) .
   \label{equ:mcorr_2}
\end{split}
\end{equation}

This final formulation of $M_\mathrm{\alpha}$ only requires the radial profile of $\alpha$ in addition to the other quantities used to estimate $M_\text{HE}$.

\subsection{The effective mass terms correction}

The second approach accounts for velocity as additional effective-mass terms. More precisely, several works \citep[e.g.][]{2009ApJ...705.1129Lau,2013ApJ...777..151Lau,2013ApJ...767...79Suto} proposed to account for the residual gas motion in clusters by using the Euler momentum equation, similarly to the Jeans equations for collisionless systems (see \citeauthor{binney2008galactic} \citeyear{binney2008galactic} for more details), called the "averaging" method. By injecting the Euler momentum flux tensor into the momentum conservation equation (see \citeauthor{2013ApJ...777..151Lau} \citeyear{2013ApJ...777..151Lau} for details), several effective mass terms are derived, accounting for the gas random motion, streaming, acceleration, bulk rotation \citep[recently investigated, e.g., in X-ray observed massive galaxy clusters by][]{2025A&A...697A..17Bartalesi}, and an additional cross term. Since this work aims to compare the 3D mass terms to their counterpart derived from projected quantities, we here only compute the random motion and bulk rotation terms, which are not only the dominant ones \citep[see e.g.][]{2013ApJ...777..151Lau,nelson2014hydrodynamic} but also the only ones accessible from the line shift and broadening measured in X-ray spectroscopy. First, the acceleration term involves the time derivative of the mean velocity field and is therefore inaccessible from observations. Second, the cross term depends on the off-diagonal components of the velocity dispersion tensor, which vanish under the isotropy assumption we adopt for the projected quantities. Finally, the streaming term requires the full 3D mean velocity field and its angular gradients, which cannot be recovered from line-of-sight projected quantities. The velocity dispersion and bulk rotation effective mass terms can be written as follows

\begin{equation}
    M_{\mathrm{disp}} = -\frac{r^2}{G \rho_{\mathrm{gas}}} \frac{d(\rho_{\mathrm{gas}} \sigma_{\rm r}^2)}{dr} -\frac{r}{G}(2\sigma_{\rm r}^2-\sigma_{\rm t}^2)
    \label{eq:m_disp}
\end{equation}

\begin{equation}
    M_{\mathrm{rot}}=\frac{r v_t^2}{G}
    \label{eq:m_rot}    
\end{equation}

with $\sigma_{\rm r}$ the velocity dispersion of the radial velocity component, $v_r$, in spherical coordinates, defined as 

\begin{equation}
    \sigma_{\rm r}^2 = \bar{v_r^2} - (\bar{v_r})^2 
\end{equation}

\noindent where the bar stands for the mass-weighted mean within a spherical shell. $v_t$ is the tangential velocity defined as $v_t^2=v_\theta^2+v_\phi^2$, also in spherical coordinates, its associated velocity dispersion being defined as $\sigma_{\rm t}^2=\sigma_{\rm \theta}^2+\sigma_{\rm \phi}^2$. Assuming once again a perfect gas and log derivatives, Eq.~\ref{eq:m_disp} then rewrites 

\begin{equation}
    M_{\mathrm{disp}} = -\frac{r}{G} \left[\sigma_{\rm r}^2\left(\frac{d\ln(n_e)}{d\ln(r)}+\frac{d\ln(\sigma_{\rm r}^2)}{d\ln(r)}+2\right) -\sigma_{\rm t}^2\right] .
    \label{eq:m_disp_log}
\end{equation}

\noindent The mass associated with these effective mass terms is thus $M_{\mathrm{eff}}=M_{\mathrm{HE}}+M_{\mathrm{disp}}+M_{\mathrm{rot}}$.

\section{Mass corrections from 3D radial profiles}
\label{sec4: 3d mass}
 We present the masses derived from 3D radial profiles and their derivatives, which give access to the true, unprojected quantities and thus serve as a reference for the projected estimates presented in Sect.~\ref{sec:2d masses}. The profiles were computed as the mass-weighted mean of each quantity in spherical shells in 20 bins of 0.1 length in logarithmic scale, ranging from 56~kpc to 4.466~Mpc, similarly to \citet{2024A&A...682A.157Lebeau}. This Virgo numerical replica shows complex gas dynamics due to its formation history, including a late merger in the simulation, and due to its local environment, as shown in \citet{2024A&A...689A..19Lebeau,2025A&A...704A..14Lebeau}. We explore the impact of these dynamics on mass estimates by computing them across four regions. Using the knowledge acquired in the previous studies of this simulation \citep[][]{2024A&A...689A..19Lebeau,2024A&A...682A.157Lebeau,2025A&A...704A..14Lebeau,2026A&A...707A.336Lebeau}, we divide the cluster into regions of equal volume, named afterwards Outflow, Collapse, Collimated filament, and Multistream filament. These regions are shown in Fig.~\ref{app:sectors_ne}, and are displayed in red, purple, green, and blue, respectively, also in the following figures. In addition, the masses are also estimated using the entire cluster; they are labeled as ``Entire cluster'' and displayed in black throughout this work. Moreover, the characteristic radii $R_\mathrm{500}$=1087~kpc and $R_\mathrm{vir}$=2147~kpc are shown as grey dotted vertical lines in all the figures. 

\begin{figure}
    \centering
    \includegraphics[trim=20 15 40 40,clip,width=1\linewidth]{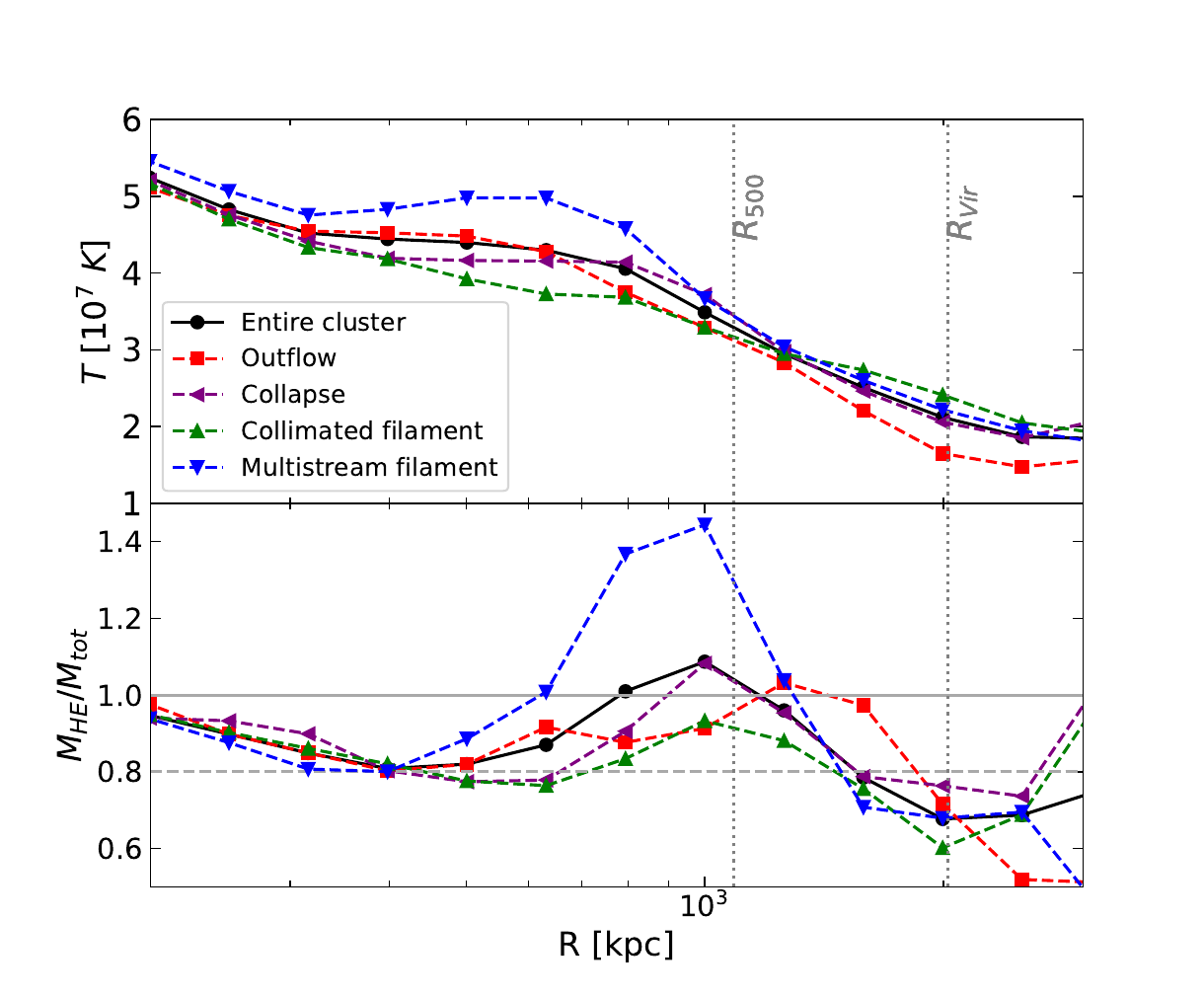}
    \caption{Top: 3D radial profiles of the temperature. Bottom: ratios of $M_{\rm HE}$ over $M_{\rm tot}=M_{\rm DM}+M_{\rm gas}$, which is the sum of the mass of all the DM particles and the gas cells in the simulation.}
    \label{T-mhe}
\end{figure}

We first present the temperature and the hydrostatic mass bias $(1-b)=M_\mathrm{HE}/M_\mathrm{tot}$ in the top and bottom panels of Fig.~\ref{T-mhe}. The total hydrostatic mass bias lies between 0.8 and 1 for $R<1500$~kpc, which is within the scatter of the value expected from other simulations \citep[see e.g.][]{gianfagna2021exploring}. These values are lower than those presented in \citet{2024A&A...682A.157Lebeau}, for which $T/\mu$ was wrongly used instead of $T$, it has been corrected in this work\footnote{Although the values of the hydrostatic mass are incorrect in \citet{2024A&A...682A.157Lebeau}, the analysis of the impact of the projection effects remains valid.}. The estimated hydrostatic mass is very sensitive to variations in the temperature radial profiles, the pressure and electron density profiles being much smoother (see Fig.~\ref{app:p_ne_3D_profs}). For instance, there is a strong accretion shock deep in the cluster, where the gas is funnelled by the filament, and already identified with the projected pressure in the top-right panel of Fig. 2 in \citet{2024A&A...682A.157Lebeau}, leading to a temperature jump at $R\sim850$~kpc and an increased temperature at lower radii in the Multistream filament region. This shock leads to an overestimation of the mass by slightly more than 40\% with respect to $M_\mathrm{tot}$ in that region. This strong temperature gradient in that region also affects the temperature profile of the entire cluster and thus its hydrostatic mass estimation, explaining why the hydrostatic mass bias is above one at this radius, though the effect is mitigated by the contribution from regions not experiencing such a shock. 

\begin{figure}
    \centering
    \includegraphics[trim=20 10 50 30,clip,width=1\linewidth]{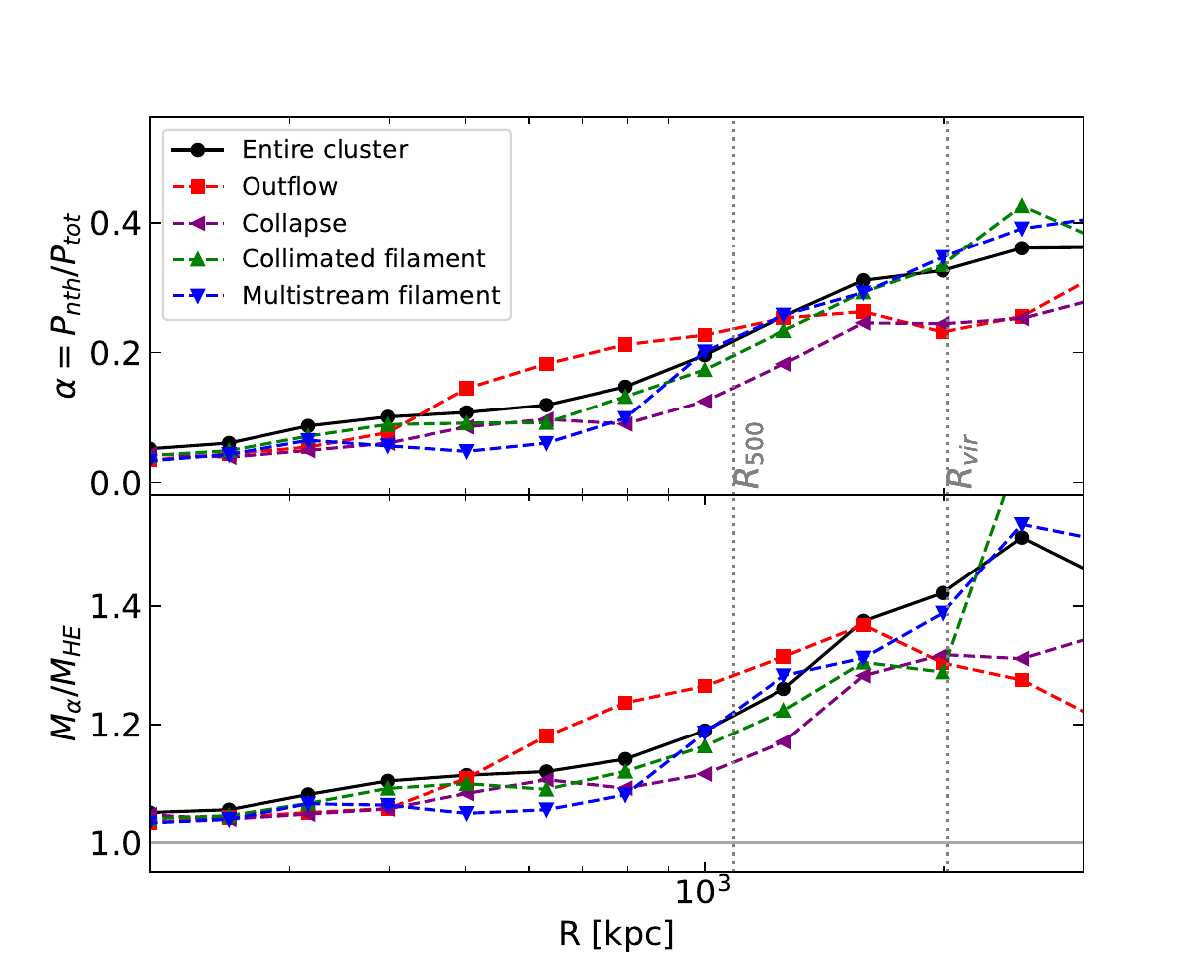}
    \caption{Top: 3D radial profiles of the $\alpha$ parameter. Bottom: ratios of $M_{\rm \alpha}$ over $M_{\rm HE}$.}
    \label{alpha-malpha}
\end{figure}

We now estimate the 3D non-thermal pressure as $P_{\rm nth}=(\rho_{\rm gas}\sigma_{\rm 3D}^2)/3$ with $\sigma_{\rm 3D}^2=\sigma_{\rm x}^2+\sigma_{\rm y}^2+\sigma_{\rm z}^2$ the 3D velocity dispersion and $\sigma_{\rm i}^2=\bar{v_{\rm i}^2}-(\bar{v_{\rm i}})^2$ the velocity dispersion for each velocity component with the bar standing for the mean. The radial profile of $P_{\rm nth}$ is computed using the same binning and mass-weighted mean method as for the other radial profiles. We directly present the radial profile of the non-thermal pressure fraction $\alpha=P_{\rm nth}/P_{\rm tot}$, with $P_{\rm tot}=P_{\rm nth}+P_{\rm th}$ the total pressure, in the top panel of Fig.~\ref{alpha-malpha}. The non-thermal pressure fraction increases with radius, as expected and consistent with other works (see e.g. Fig.~10 of \citeauthor{gianfagna2021exploring} \citeyear{gianfagna2021exploring}). For the entire cluster, $\alpha$ is below 0.1 in the core, reaches $\sim$0.22 at $R_{\rm 500}$, and increases up to $\sim$0.33 at $R_{\rm vir}$. 

The Outflow region (red) has the highest non-thermal pressure fraction in the [500,1000]~kpc range, with $\alpha$ rising from 0.11 at 500~kpc to 0.26 at 1000~kpc, which can be explained by the outflow encountering the accreted gas at this radius, generating shocks, mixing, and non-thermal gas motion (see the bottom-right part of the left panel of Fig.~2 in \citeauthor{2025A&A...704A..14Lebeau} \citeyear{2025A&A...704A..14Lebeau}). Beyond $R_{\rm vir}$, $\alpha$ rises above 0.3 in the Collimated and Multistream filament regions and for the entire cluster, whereas it flattens in the Outflow and Collapse regions. This is due to turbulent motions in the filaments, whereas the accretion flows in the other regions are laminar, thus with a smaller velocity dispersion, as we showed in \citet{2025A&A...704A..14Lebeau}. The increase of $\alpha$ with radius induces an increase of its contribution to the corrected mass, $M_{\alpha}$, with respect to the hydrostatic mass as shown in the bottom panel of Fig.~\ref{alpha-malpha}. The corrected mass is about 5\% higher than $M_{\rm HE}$ in the core, about 20\% higher at $R_{\rm 500}$, and around 40\% at $R_{\rm vir}$. We discuss the impact on the hydrostatic mass bias and compare these results with the second correction method at the end of this section.

\begin{figure}
    \centering
    \includegraphics[trim= 0 0 0 0,clip,width=1\linewidth]{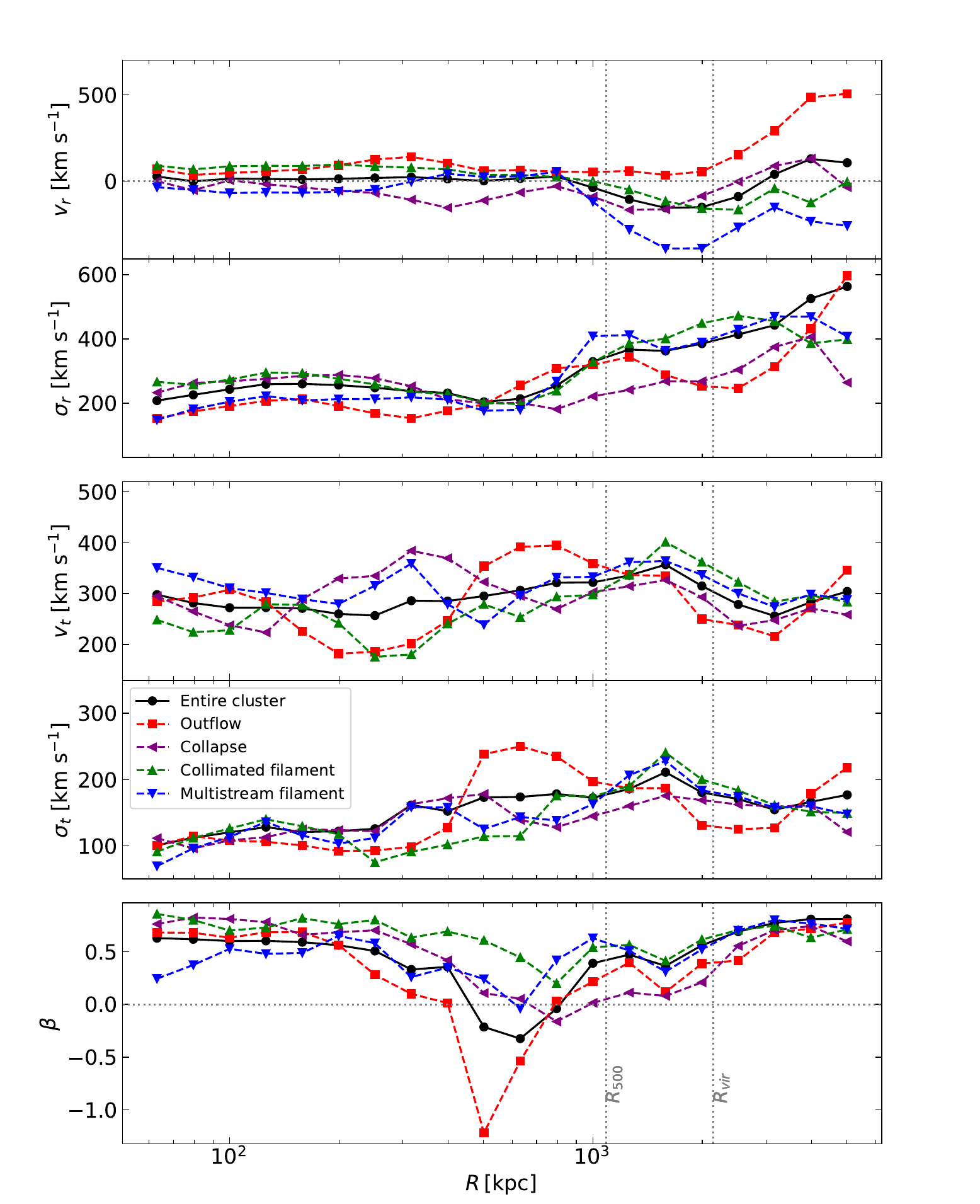}
    \caption{3D radial profiles of the radial velocity (top row), its dispersion (second row), the tangential velocity (third row), its dispersion (fourth row), and the anisotropy parameter $\beta = 1 - \sigma_t^2/2\sigma_r^2$ (bottom row).}
    \label{vel_rad_prof_3D}
\end{figure}

Before presenting the estimation of the effective mass terms, we show the radial profiles of the radial and tangential velocity components and their dispersion in Fig.~\ref{vel_rad_prof_3D}. Within 900~kpc, the mean radial velocity is close to $0~\mathrm{km\,s^{-1}}$ in all regions, indicating that the Virgo replica is virialised at these radii. Beyond 900~kpc, the radial velocity shows a large scatter among regions: it drops below $-300~\mathrm{km\,s^{-1}}$ in the Multistream filament region, indicating strong mass accretion, and is highly positive in the Outflow region. The radial velocity dispersion follows a similar trend, increasing from $\sim 200~\mathrm{km\,s^{-1}}$ within 900~kpc to almost $600~\mathrm{km\,s^{-1}}$ in the filament outskirts. In contrast, the tangential velocity and its dispersion are more uniform with radius. The gas is clearly not at equilibrium in the [900,2000]~kpc range, which includes $R_{\mathrm{500}}$ and $R_{\mathrm{vir}}$, confirming the necessity to account for the out-of-equilibrium gas motions at the radii where the cluster mass is typically estimated. The departure from isotropy can be quantified through the anisotropy parameter \citep[see e.g.][]{nelson2014hydrodynamic,shi2016locations}
\begin{equation}
    \beta = 1 - \frac{\sigma_{\rm t}^2}{2\sigma_{\rm r}^2} ,
    \label{eq:beta}
\end{equation}
\noindent with $\sigma_{\rm t}^2 = \sigma_{\rm \theta}^2 + \sigma_{\rm \phi}^2$, which is zero for an isotropic velocity dispersion tensor, positive for radially-biased motions, and negative for tangentially-biased ones. As shown in the bottom panel of Fig.~\ref{vel_rad_prof_3D}, $\beta$ is positive across most of the radial range for the entire cluster, with a value of $\sim$0.6 in the core, indicating a velocity tensor biased towards radial motions, consistent with the dominant accretion from filaments. The notable exception is the [400--900]~kpc range in the Outflow region, where $\beta$ drops well below zero, reflecting the tangential nature of the outflow bubble identified in the $v_{\rm t}$ and $\sigma_{\rm t}$ profiles. The anisotropy parameter thus departs substantially from zero across the entire cluster, confirming that the isotropy assumption necessarily adopted for the projected estimates in Sect.~\ref{sec:2d masses} is a strong simplification of the velocity field in this Virgo replica. Moreover, this behaviour of the $\beta$ parameter is not in agreement with the average value at small radii of the sample of \citet{nelson2014hydrodynamic} (see Fig. 4 of their work). In their case, $\beta$ is around 0.2 for $R<0.4\times R_{\rm 200c}$, corresponding to R<700~kpc for our Virgo replica\footnote{$R_{\rm 200c}=1.07$~Mpc for this Virgo replica, see \citet{2024A&A...689A..19Lebeau}}, whereas it is above 0.4 in ours. However, the agreement is better at larger radii. This difference emphasises how peculiar the core dynamics of this Virgo replica is. 

\begin{figure}
    \centering
    \includegraphics[trim= 35 60 60 60,clip,width=1\linewidth]{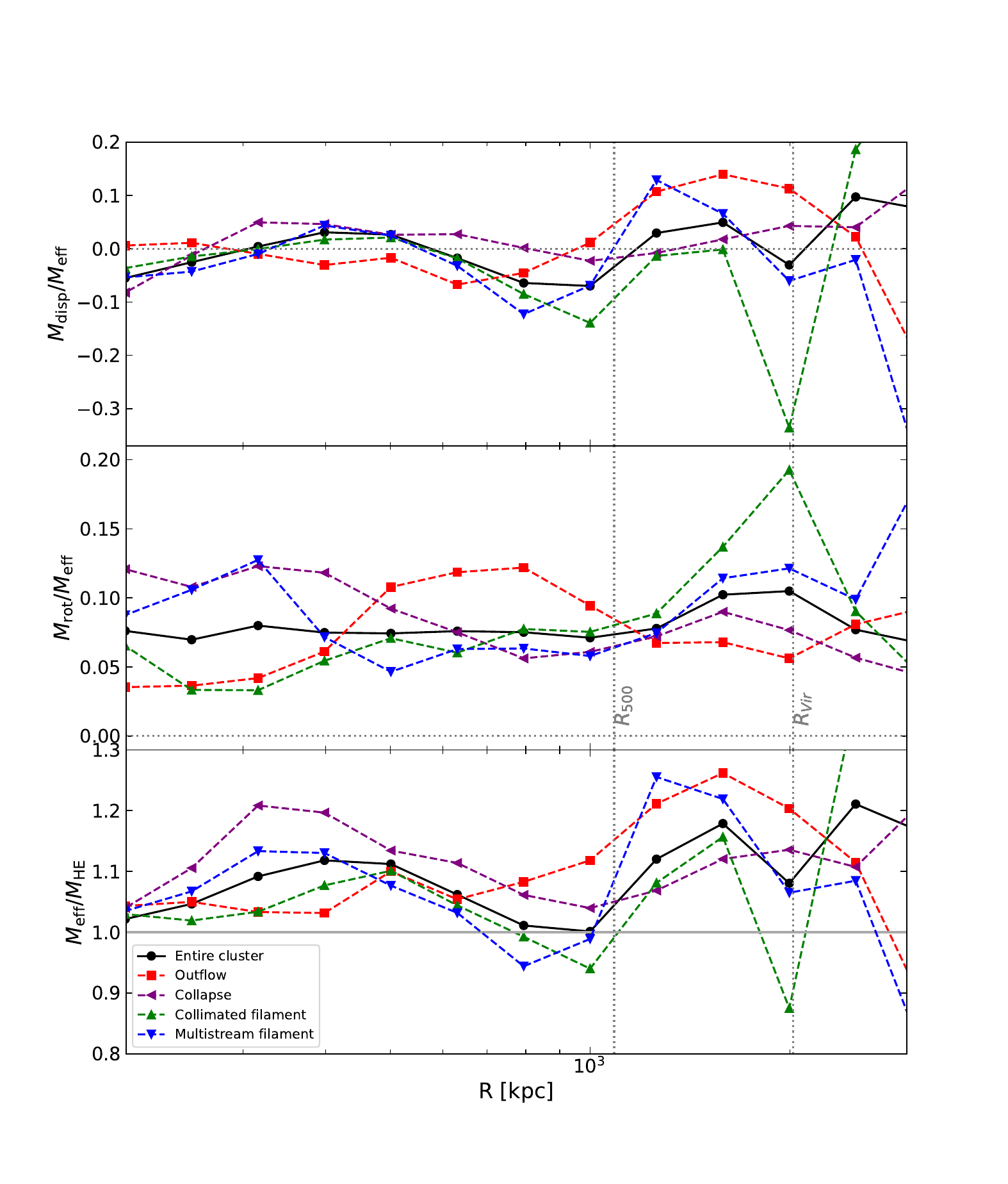}
    \caption{Ratios of $M_{\rm disp}$ (top), $M_{\rm rot}$ (middle) over $M_{\rm eff}=M_{\rm HE}+M_{\rm disp}+M_{\rm rot}$, and $M_{\rm eff}$ over $M_{\rm HE}$ (bottom).}
    \label{Mdisp_Mrot_MJ_3D}
\end{figure}

We now use Eqs.~\ref{eq:m_disp} and~\ref{eq:m_rot} to compute the effective mass terms. Their relative contributions to $M_{\mathrm{eff}}$ are shown in Fig.~\ref{Mdisp_Mrot_MJ_3D}. The contribution of $M_{\mathrm{disp}}$ is contained within [-0.15,0.05] inside $R_{\mathrm{500}}$ and shows no clear trend with the dynamics in the different regions: as shown in Eq.~\ref{eq:m_disp_log}, it depends on both $\sigma_{\rm r}^2$ and $\sigma_{\rm t}^2$ and on the gradient of $\sigma_{\rm r}^2$, which can have opposite signs and partially cancel out. In contrast, $M_{\mathrm{rot}}$ is positive by definition and contributes between 3 and 14\% to $M_{\mathrm{eff}}$ within $R_{\mathrm{500}}$, with a roughly constant $\sim$7\% for the entire cluster. Beyond $R_{\mathrm{500}}$, the contributions are of similar magnitude in most regions. As shown in the bottom panel, comparing $M_{\mathrm{eff}}$ to $M_{\mathrm{HE}}$, the contribution of these two effective mass terms increases the total estimated mass at most radii, by a few to about 25\% with respect to $M_{\mathrm{HE}}$, regardless of the region. Overall, the gas rotation needs to be accounted for in this framework.

To conclude this section, we compare the mass biases from the two correction methods in Fig.~\ref{Mtot_comp_3D} (see also Fig.~\ref{app:M_ratio_comp_3D} in Appendix~\ref{app:sec:mass comp}). For $R<600$~kpc, $M_{\rm eff}$ is slightly higher than $M_{\alpha}$, but the trend reverses at larger radii, as expected given the increasing $M_{\alpha}/M_{\rm HE}$ ratio with radius compared to the less varying $M_{\rm eff}/M_{\rm HE}$. For the entire cluster, the two corrections are nearly indistinguishable for $R < 600$~kpc, with the difference smaller than the bin-to-bin fluctuations induced by the numerical derivatives of the radial profiles. The difference between the two corrections becomes significant only beyond $R_{500}$, where the rising $\alpha$ drives $M_{\alpha}$ above $M_{\rm eff}$. Both corrections reduce the mass bias by a few per cent for $R<600$~kpc. However, where the hydrostatic mass already overestimates $M_{\rm tot}$, particularly in the [800--1500]~kpc range in the Multistream filament region, the positive corrections of both approaches further increase the bias. Overall, the two correction methods do not yield the same results across all radii and regions, which we discuss further in Sect.~\ref{discussion}.

\begin{figure}
    \centering
    \includegraphics[trim= 25 65 50 100,clip,width=1\linewidth]{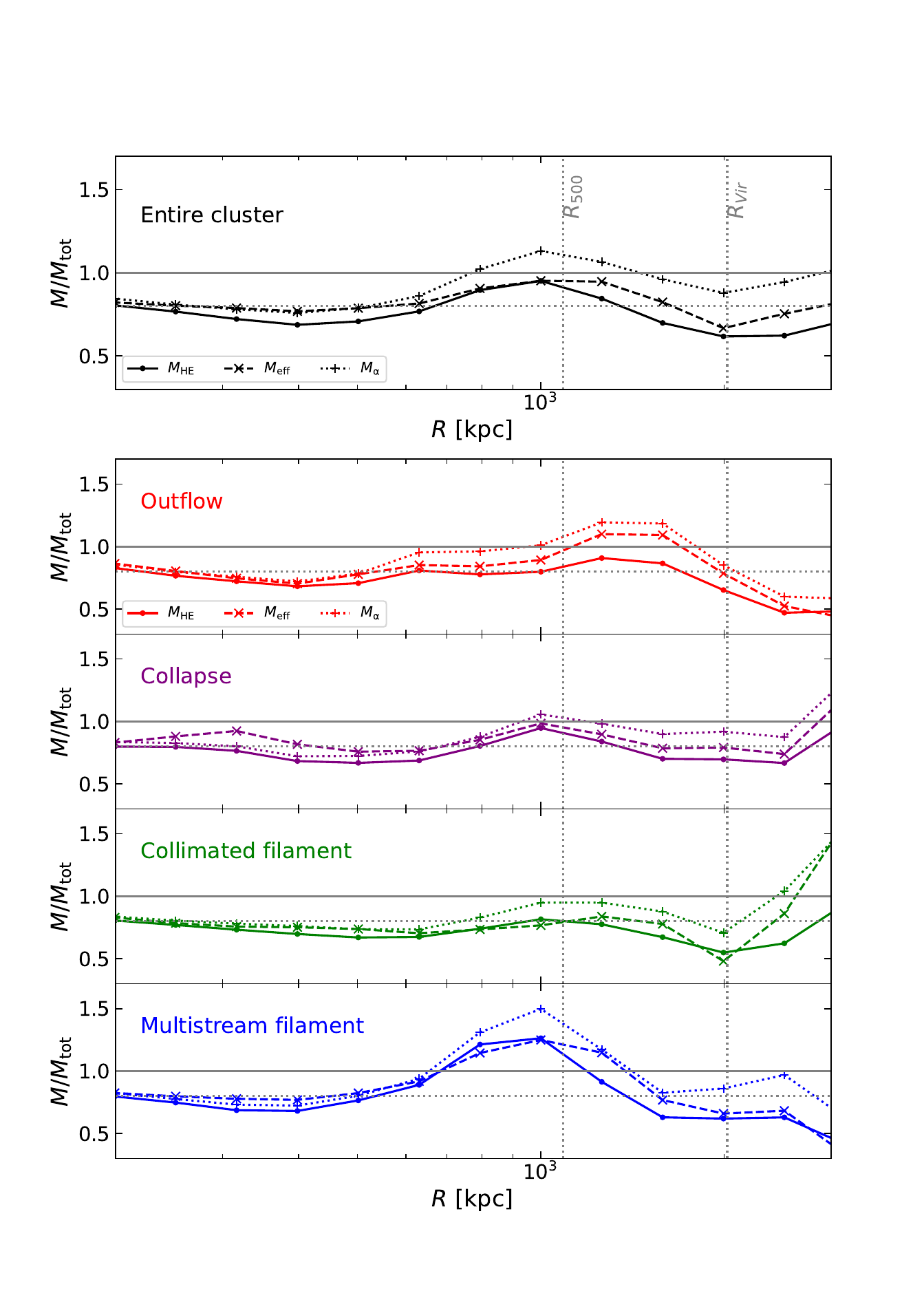}
    \caption{Ratios of $M_{\rm HE}$ (solid), $M_{\rm eff}$ (dashed) and $M_{\rm \alpha}$ (dotted) over $M_{\rm tot}$ using all Virgo's regions (top row in black), in the Outflow region (second row in red), in the Collapse region (third row in purple), in the Collimated filament region (fourth row in green), and in the Multistream filament region (bottom row in blue).}
    \label{Mtot_comp_3D}
\end{figure}

\section{Masses from projected quantities}
\label{sec:2d masses}

After computing the corrected masses using 3D radial profiles in different regions of the Virgo replica, we now estimate these masses using projected quantities, as is done in observations. However, it is worth noting that the projections used are not mock observations; they are projections of the velocity and its dispersion along sightlines made to mimic the quantities derived from the line shift and broadening in high-resolution X-ray spectroscopy, as currently done with XRISM \citep[e.g.][]{2025PASJ...77S.242XRISMA2029} and in the future with NewAthena \citep[see e.g.][]{roncarelli2018athena}. We defer the generation of mock observations to a future article.

More precisely, we project the velocity and compute its dispersion along four sightlines: the \textit{x}, \textit{y}, and \textit{z} axes of the simulation box, and along the sightline between the center of the Local Universe simulation box and the center of the Virgo cluster, to mimic our observer's sightline (labeled cen hereafter). Two weighting schemes are used: emission- and mass-weighted, denoted ew and mw, respectively. The details of the projection procedure and the resulting maps can be found in \citet{2026A&A...707A.336Lebeau}.

Then, to mimic what could be observed by XRISM, that is, a high spectral resolution but a limited spatial resolution, we do not compute the radial profiles of the sightline velocity and its dispersion; we rather use their mean value within $R_{\rm 500}$ and $R_{\rm vir}$. We thus cannot estimate the log-derivative of $\sigma_{\rm r}^2$ in Eq.~\ref{eq:m_disp_log}, this term is thus set to zero. Moreover, we make the usual assumption that the ICM gas motion is isotropic since we only have access to 1D information through spectroscopy, even though the anisotropy parameter $\beta$ (Eq.~\ref{eq:beta}, see bottom panel of Fig.~\ref{vel_rad_prof_3D}) shows that this is not the case in this Virgo replica, as also discussed in \citet{2026A&A...707A.336Lebeau}. We thus assume that $\sigma_{\rm 1D}^2=\sigma_{\rm los}^2=\sigma_{\rm 3D}^2/3$ (los=line-of-sight). Consequently, the non-thermal pressure is calculated as $P_\text{nth}=(\rho\sigma_{\rm los}^2)$ from the projections. The hypothesis of isotropy also implies that $\sigma_{\rm r}^2=\sigma_{\rm \theta}^2=\sigma_{\rm \phi}^2=\sigma_{\rm t}^2/2$, thus simplifying the velocity dispersion effective mass term of Eq.~\ref{eq:m_disp_log} into 

\begin{equation}
    M_{\mathrm{disp}} = -\frac{r}{G}\sigma_{\rm r}^2\frac{d\ln(n_e)}{d\ln(r)} .
    \label{eq:m_disp_log_1d}
\end{equation}

\noindent It is important to note that this adapted mass term is, by definition, positive, since the radial gradient of the electron density is negative in the ICM, whereas its 3D version can take negative values. On the other hand, $M_{\rm rot}$ remains as in Eq.~\ref{eq:m_rot}, $v_{\rm t}$ is estimated as the maximum velocity difference between opposite quadrants on the projections, as detailed in Appendix \ref{app:v_t}. 

We now present the results in the same order as in the previous section, that is, (i) the non-thermal pressure mass correction, (ii) the effective mass terms, and (iii) a comparison of the approaches and their impact on the mass bias. In the following figures, the \textit{x}, \textit{y}, \textit{z} and \textit{cen} projections are displayed respectively in orange, blue, red and pink with crosses (dots) for the ew (mw) projections. In addition, the 3D radial profile of each quantity computed using the entire cluster is displayed as a black line surrounded by a grey shaded area representing the standard deviation among the five values estimated for the entire cluster and within the four regions studied in the 3D case above. Here, we focus on the mass added by these correction methods, considering ratios as in the previous section; the mass values are shown in Tab.~\ref{app: masses tab}.

\begin{figure}
    \centering
    \includegraphics[trim= 15 25 40 70,clip,width=1\linewidth]{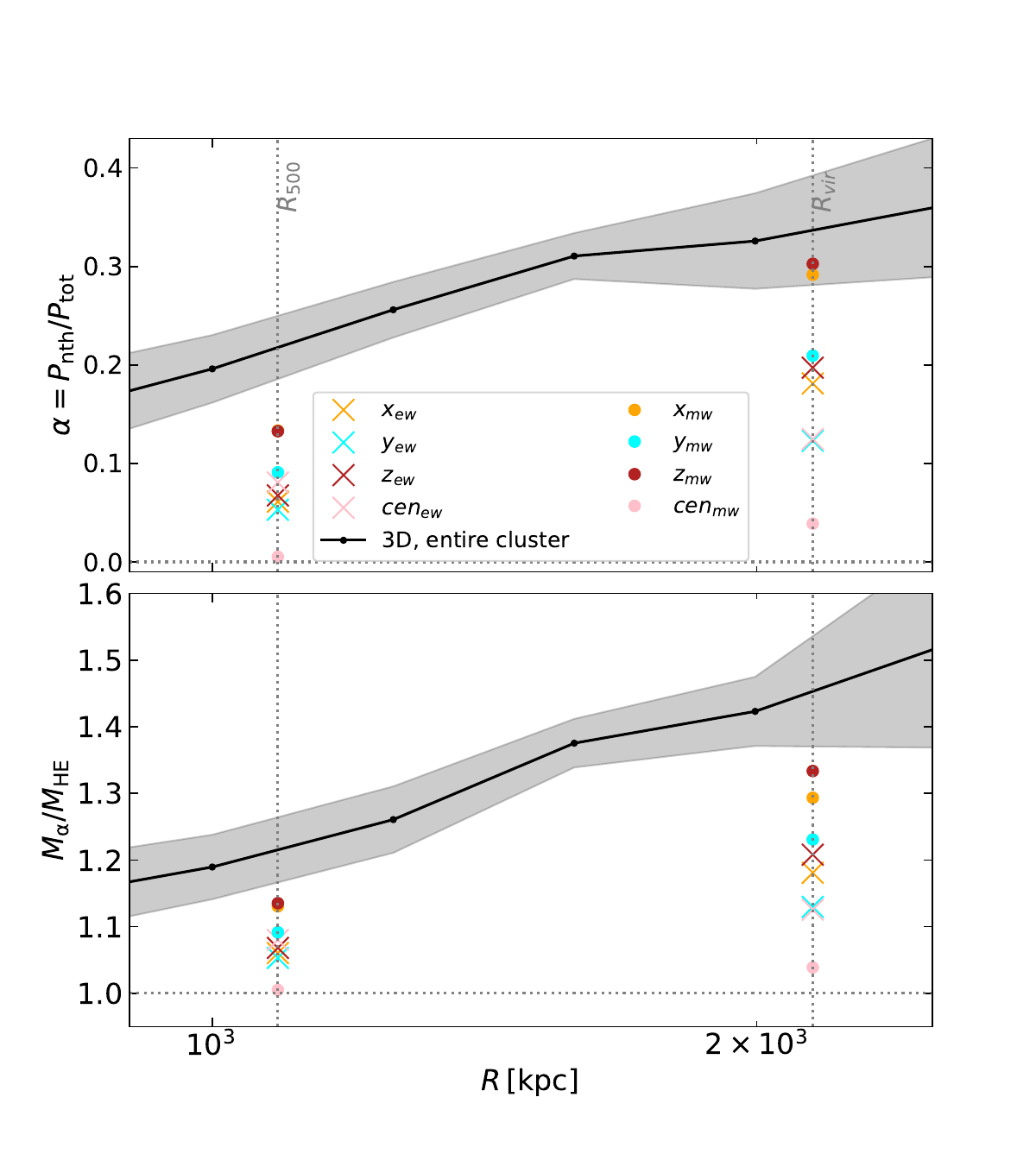}
    \caption{Values of the $\alpha$ parameter (top) and the $M_{\rm \alpha}/M_{\rm HE}$ ratio estimated within $R_{\rm 500}$ and $R_{\rm vir}$ from projections.}
    \label{alpha-malpha_2d}
\end{figure}

First, in the top panel of Fig.~\ref{alpha-malpha_2d}, we present the non-thermal pressure fraction estimated at $R_{\mathrm{500}}$ and $R_{\mathrm{vir}}$ from projections. At $R_{\mathrm{500}}$, most values are in the [0.05,0.1] range, about a factor of two lower than the 3D value of $\sim$0.22. The three outliers are all mass-weighted projections. For $x_{\mathrm{mw}}$ and $z_{\mathrm{mw}}$, filaments in Virgo's local environment produce a bimodal sightline velocity distribution (see Fig.~1 in \citeauthor{2026A&A...707A.336Lebeau} \citeyear{2026A&A...707A.336Lebeau}), increasing the velocity dispersion and thus $\alpha$, an effect amplified by the mw scheme. Conversely, for $cen_{\mathrm{mw}}$, the sightline is aligned with the filaments connected to Virgo, so the velocity field is dominated by the coherent accretion flow, yielding a low dispersion and thus a low $\alpha$, again amplified by the mw scheme. Similar trends are observed at $R_{\mathrm{vir}}$, where unbiased projections scatter in the [0.10,0.20] range compared to $\sim$0.33 for the 3D case.

These results directly transfer to the $M_{\alpha}/M_{\mathrm{HE}}$ ratio (bottom panel). At $R_{\mathrm{500}}$, $M_{\alpha}$ increases by a few to 10\% for the unbiased projections, compared to $\sim$20\% in 3D. At $R_{\mathrm{vir}}$, the increase is scattered around 20\% from projections, compared to $\sim$40\% in 3D. The projected values are thus approximately a factor of two lower than the 3D ones.

\begin{figure}
    \centering
    \includegraphics[trim= 15 50 40 100,clip,width=1\linewidth]{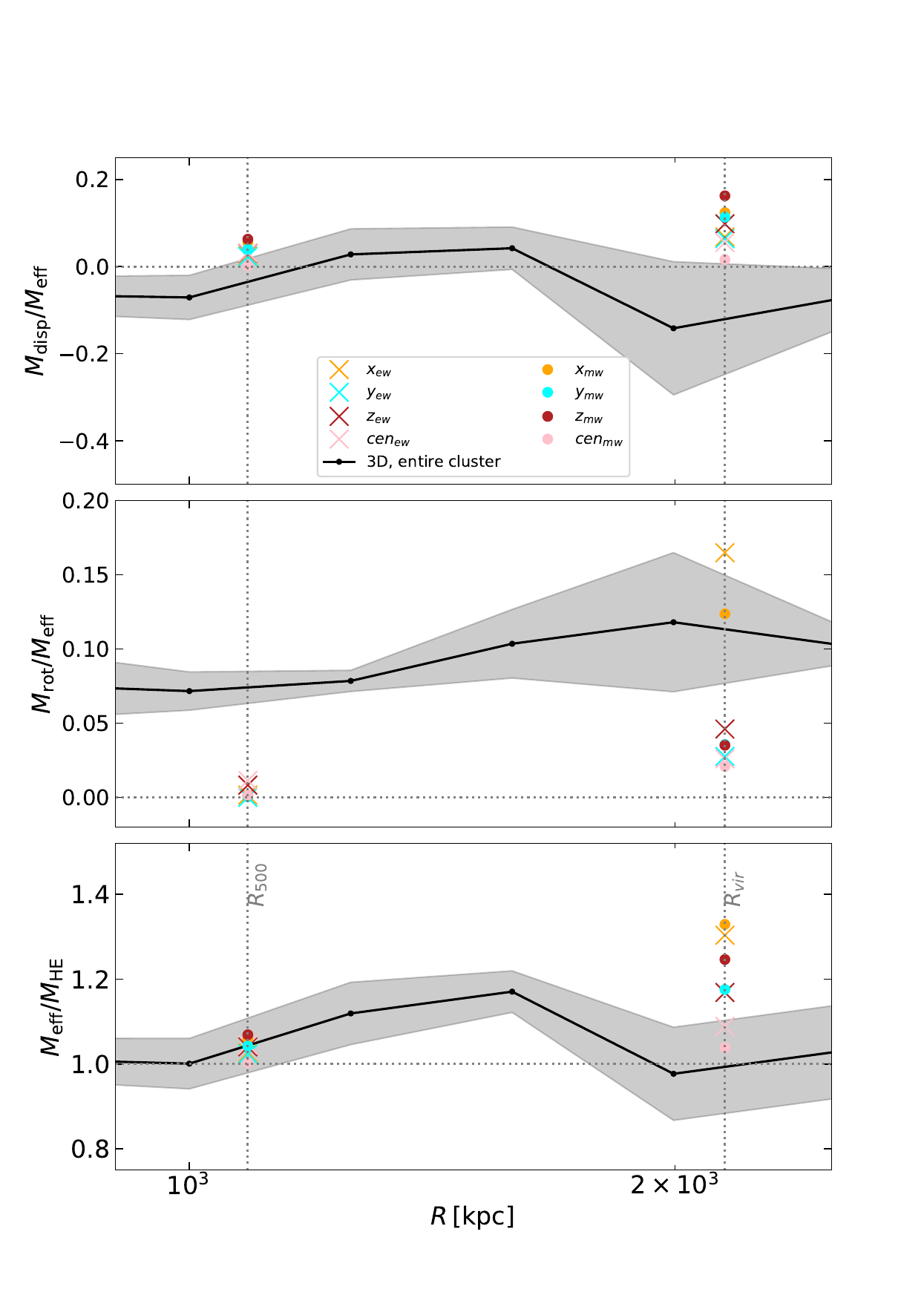}
    \caption{Ratios of $M_{\rm disp}$ (top), $M_{\rm rot}$ (middle) over $M_{\rm eff}=M_{\rm HE}+M_{\rm disp}+M_{\rm rot}$, and $M_{\rm eff}$ over $M_{\rm HE}$ (bottom) estimated from projections.}
    \label{Mdisp_Mrot_MJ_2D}
\end{figure}

Then, the relative contributions of $M_{\rm disp}$ and $M_{\rm rot}$ to $M_{\rm eff}$, and the $M_{\rm eff}/M_{\rm HE}$ ratio, are shown in the top, middle and bottom panels of Fig.~\ref{Mdisp_Mrot_MJ_2D}, similarly to Fig.~\ref{Mdisp_Mrot_MJ_3D}. At $R_{\rm 500}$, $M_{\rm disp}$ contributes to $M_{\rm eff}$ in the [2,6]\% range, which is higher than the negative value in the 3D case, though their absolute values are rather close. At $R_{\rm vir}$, the ratios are in the [0.01,0.16] range, which is scattered around the absolute value of the 3D entire cluster case. In summary, the contributions of $M_{\rm disp}$ to $M_{\rm eff}$ estimated from projections, which are positive by definition, are less scattered than the 3D values and weakly compatible with the latter. The same distribution of values among the projections appears as in the previous figure, with, for instance, the highest being that of the $z_{\rm mw}$ projection and the lowest being that of the $cen_{\rm mw}$ projection, due to the same projection effects as above.

The middle panel shows that $M_{\rm rot}$ has a much lower contribution to $M_{\rm eff}$ when estimated from projections than in the 3D case, with less than 2\% at $R_{\rm 500}$ and 5\% at best at $R_{\rm vir}$. This is expected, as a coherent rotation pattern is largely smeared out when integrated along the line of sight, since regions of opposite line-of-sight velocity overlap in projection and partially cancel. In addition, the quadrant estimator of \citet{2018PASJ...70...51Ota} only recovers the residual large-scale velocity difference between opposite quadrants, which is a lower limit on the true bulk rotation. To our knowledge, no estimator allows a direct recovery of $v_{\rm t}$ from projected data, and improving it is beyond the scope of this work. Both effects, the smearing along the line of sight and the limitation of the estimator, explain why $M_{\rm rot}$ is systematically lower in projection than in 3D. The only exception is the x projection, both mw and ew, which is within the dispersion of the 3D case due to the bipolar velocity distribution induced by the filaments connected to the cluster, as discussed above.

Finally, in the bottom panel, we can see that, at $R_{\rm 500}$, though the contribution from $M_{\rm disp}$ ($M_{\rm rot}$) is higher (lower) than in the 3D case, the mass added to $M_{\rm eff}$ by these two effective mass terms is unexpectedly in good agreement with the values found in the 3D case, that is about 5\%. This may simply be because the contributions are minor in both cases. At $R_{\rm vir}$, however, the mass added by these two terms is in the [17,33] per cent range from projections, whereas it is in the [-16,17] per cent range in the 3D cases. This is because the two terms compensate each other in the 3D cases (see Tab.~\ref{app: masses tab}), whereas they add up in projections since $M_{\rm disp}$ can only be positive. Again, the \textit{x} and \textit{z} projections substantially increase the mass estimation at this radius.

\begin{figure}
    \centering
    \includegraphics[trim= 20 40 20 80,clip,width=1\linewidth]{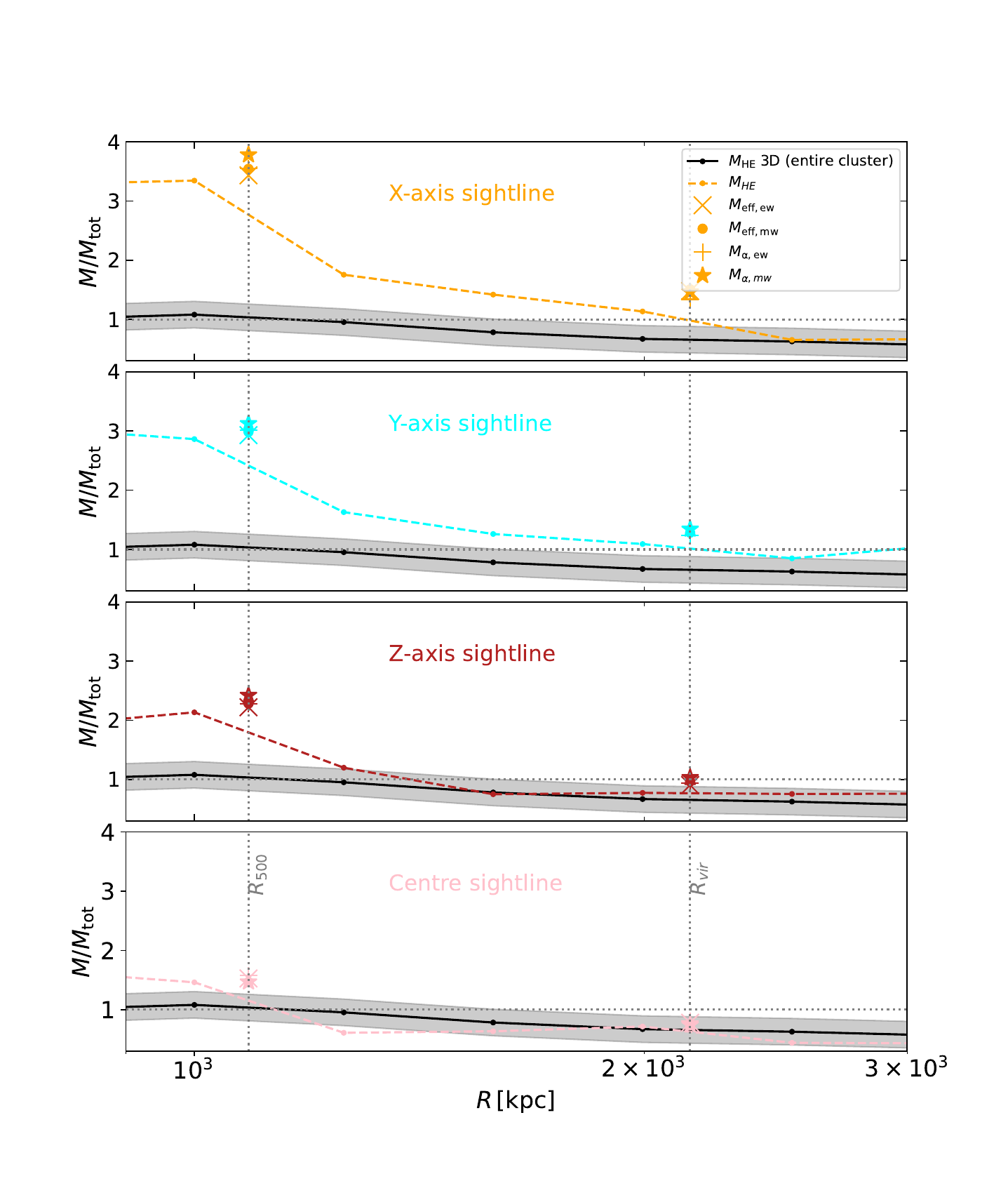}
    \caption{Ratios of $M_{\rm HE}$ (dashed line), $M_{\rm eff,ew}$ (cross), $M_{\rm eff,mw}$ (dot), $M_{\rm \alpha,ew}$ (plus) and $M_{\rm \alpha,mw}$ (star) over $M_{\rm tot}$ estimated from projections along the X-axis (top row in orange), the Y-axis (second row in blue), the Z-axis (third row in red) and the Centre sightlines (bottom row in pink).}
    \label{Mtot_comp_2D}
\end{figure}

We finally compare the mass biases estimated from the two correction methods for the four sightlines in Fig.~\ref{Mtot_comp_2D}. In each panel, the 3D hydrostatic mass bias is shown as a black solid line with a grey shaded area, and the deprojected hydrostatic mass bias is shown in dashed lines. The deprojected profiles are computed using the model-free geometrical method of \citet{2024A&A...682A.157Lebeau} (see Fig.~6 therein), which we showed amplifies projection effects and significantly overestimates the hydrostatic mass, in particular at $R_{\mathrm{500}}$. This limits our ability to draw conclusions about the absolute corrected mass bias from projections, but still allows a comparison between the two approaches.

At both $R_{\mathrm{500}}$ and $R_{\mathrm{vir}}$, and for both weighting schemes, the two correction approaches are in good agreement with each other for a given sightline. At $R_{\mathrm{500}}$, the corrections add a few per cent of mass in all sightlines, but since the hydrostatic mass is already overestimated, this further increases the bias. At $R_{\mathrm{vir}}$, the results depend on the sightline. In the \textit{z}-sightline, the corrected mass bias is close to unity for all methods and weighting schemes because the velocity distribution is the closest to Gaussian among all the sightlines \citep{2026A&A...707A.336Lebeau}, and thus the closest to the idealised case of a relaxed cluster with isotropic motions. This sightline is almost perpendicular to the two filaments \citep[see][]{2024A&A...682A.157Lebeau} and is therefore not affected by their flows. In contrast, in the \textit{cen}-sightline, which is aligned with the filaments connected to Virgo, the corrections have little impact because the sightline velocity is dominated by the gas accretion flow, leading to a low velocity dispersion and thus a small correction.

\section{Discussion}
\label{discussion}

\subsection{Comparison of the two correction approaches}

We have shown that the two approaches investigated in this work, namely the non-thermal pressure correction ($M_\alpha$) and the effective mass terms correction ($M_{\rm eff}$), do not yield identical results when applied to the 3D radial profiles in the different regions of our Virgo replica. Specifically, $M_\alpha$ increasingly exceeds $M_{\rm HE}$ with radius, driven by the rise of the non-thermal pressure fraction $\alpha$ in the outskirts, whereas $M_{\rm eff}$ shows a correction that varies less with radius, with the contributions of $M_{\rm disp}$ and $M_{\rm rot}$ partially compensating each other. This difference is expected from a physical standpoint: $M_\alpha$ accounts for the overall non-thermal pressure support regardless of the nature of the underlying gas motion, whereas $M_{\rm eff}$ decomposes it into distinct kinematic components (random motion and bulk rotation), which can have opposite signs. As a consequence, $M_\alpha$ tends to overestimate the correction in the outskirts, where the non-thermal pressure fraction is high but where the effective mass terms partially cancel out. 

\begin{figure}
    \centering
    \includegraphics[trim= 20 20 40 50,clip,width=1\linewidth]{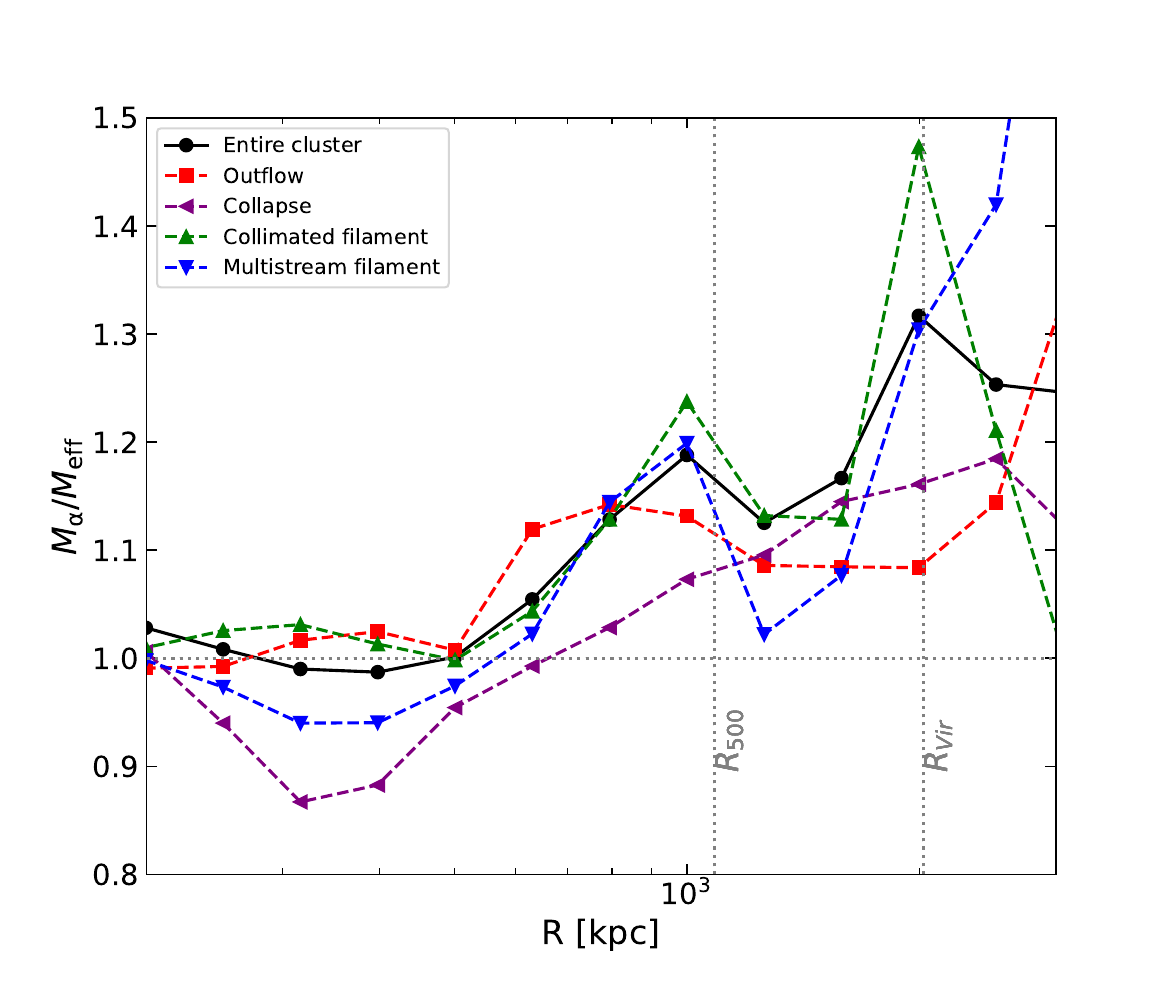}
    \caption{Ratio of $M_{\alpha}$ over $M_{\rm eff}$. The large fluctuations of the filament regions beyond $R_{\rm vir}$ are due to the steep gradients of the velocity dispersion entering the numerical derivatives in these regions.}
    \label{Malpha_Meff_ratio}
\end{figure}

Quantitatively, the ratio $M_{\alpha}/M_{\rm eff}$ is shown in Fig.~\ref{Malpha_Meff_ratio}. For the entire cluster, it remains close to unity in the core ($R<500$~kpc) and increases outwards, though not monotonically, reaching $\sim$1.13 at $R_{500}$ and $\sim$1.3 at $R_{\rm vir}$. The difference is even more pronounced in the Collimated filament regions, where the ratio rises higher than $\sim$1.45 around $R_{\rm vir}$, whereas it stays around $\sim$1.1 in the Outflow region. This confirms that the two correction approaches agree in the core but diverge increasingly in the outskirts, where the choice of correction method becomes important, particularly along the filaments.

Interestingly, when estimated from projections, the two correction methods agree much better with each other for a given sightline, both at $R_{500}$ and $R_{\rm vir}$. This is partly because the isotropy assumption imposed on the projected quantities (i.e. $\sigma_{\rm r}^2 = \sigma_{\rm \theta}^2 = \sigma_{\rm \phi}^2$, or equivalently $\beta=0$) removes the anisotropic information that distinguishes the two correction methods in 3D, despite the substantial deviation from isotropy shown in Fig.~\ref{vel_rad_prof_3D} ($\beta\sim$0.6 in the core and significantly deviating from zero across most radii). The agreement is also partly due to $M_{\rm disp}$ being positive by construction in the projected case (Eq.~\ref{eq:m_disp_log_1d}), which prevents the compensation between the two effective mass terms observed in 3D. Nevertheless, the overall agreement from projections suggests that, in practice, the choice of correction method may matter less than the accuracy of the velocity dispersion measurement since the two methods differ by less than a few per cent for a given projected sightline, whereas projection effects alone change the inferred non-thermal pressure fraction by up to a factor of two.

\subsection{Comparison with previous works and impact on the mass bias}

The non-thermal pressure fraction profiles obtained in the 3D case are in good agreement with previous simulation-based studies. In particular, the increase of $\alpha$ from below 10\% in the core to $\sim$22\% at $R_{500}$ and $\sim$33\% at $R_{\rm vir}$ is consistent with the results of \citet{nelson2014hydrodynamic}, \citet{2020MNRAS.495..864Angelinelli}, and \citet{pearce2020hydrostatic}, though with a significant scatter among the different regions, reflecting the complex and anisotropic dynamics of Virgo's local environment. The values of $\alpha$ in the filament regions rise above 0.3 beyond $R_{\rm vir}$, which is higher than the typical values found in large cluster samples but consistent with those found in disturbed systems in \citet{2020MNRAS.495..864Angelinelli}. This is likely due to the enhanced turbulence in the filaments connected to Virgo, as shown in \citet{2025A&A...704A..14Lebeau}.

We note, however, that we define the non-thermal pressure as $P_{\rm nth} = \rho \sigma_{\rm 3D}^2/3$, using the velocity dispersion computed in spherical shells without multi-scale filtering to separate bulk from turbulent motions. This is the same approach as in \citet{nelson2014hydrodynamic} and \citet{pearce2020hydrostatic}, but differs from that of \citet{2018MNRAS.481L.120Vazza} and \citet{2020MNRAS.495..864Angelinelli}, who showed that the turbulent velocity filtered from bulk motions yields lower non-thermal pressure fractions. Our values of $\alpha$ should thus be considered as upper limits on the turbulent contribution, and future comparisons using filtered velocities would help to disentangle the respective roles of bulk and turbulent motions. Similarly, the simulation does not include magnetic fields and cosmic rays, both of which could affect the dissipation of turbulent motions and the resulting non-thermal pressure support \citep[see e.g.][]{2019SSRv..215...24Simionescu}. 
These simplifications do not affect the comparison between the two correction methods, which rely on the same underlying velocity field, but they do affect the absolute values of $\alpha$ and, consequently, the corrected masses.

Regarding the analytical modeling of $\alpha$, \citet{2014MNRAS.442..521Shi} proposed a model based on the mass accretion history of clusters that predicts a radially increasing non-thermal pressure fraction, in qualitative agreement with our results. \citet{2016MNRAS.455.2936Shi} further showed that this model can be used to correct the hydrostatic mass bias, similarly to the $M_\alpha$ approach used in this work. However, their model assumes a universal profile for $\alpha$ that does not account for the anisotropic environment within individual clusters, which we have shown has a significant impact on the non-thermal pressure fraction.

The effective mass terms approach has been less widely used in the literature than the non-thermal pressure correction, but our results are qualitatively consistent with \citet{2009ApJ...705.1129Lau} and \citet{2013ApJ...777..151Lau}, who found that the velocity dispersion term can be either positive or negative depending on the radial profile of $\sigma_{\rm r}$ relative to $\sigma_{\rm t}$, and that the rotation term provides a roughly constant positive contribution. The estimation of $M_{\rm rot}$ from projections using the quadrant estimator (Appendix~\ref{app:v_t}), following \citet{2018PASJ...70...51Ota}, yields lower values than in the 3D case, suggesting that this estimator may underestimate the ICM bulk rotation. Alternative techniques for estimating $v_t$ from projected velocity maps would be worth exploring in future work.

These methodological considerations directly affect the resulting mass biases, which we provide in Tab.~\ref{app: bias tab}. The 3D hydrostatic mass bias of 0.95 at $R_{500}$ is consistent with the upper end of the $(1-b) \simeq 0.8$--$0.9$ range found in large simulation samples \citep[e.g.][]{2007ApJ...668....1Nagai,biffi2016nature,pearce2020hydrostatic,gianfagna2021exploring}, likely reflecting the specific dynamical state of this relatively low-mass and unrelaxed cluster. The overcorrection by $M_\alpha$ to 1.13 at this radius is consistent with the finding of \citet{biffi2016nature} that the object-to-object variation is large and that a correction based on $\alpha$ does not necessarily bring the bias closer to unity for individual clusters. At $R_{\rm vir}$, the 3D bias of 0.62 is significantly lower than the typical values, which is expected given the complex dynamics in the outskirts, including strong accretion flows and turbulence in the connected filaments. For the projected cases, the \textit{z}-sightline stands out as the most favorable case, with corrected biases of 0.85 ($M_{\rm \alpha,ew}$) and 0.83 ($M_{\rm eff,ew}$), closer to one than the uncorrected value of 0.71.

\subsection{From projections to observations}

The comparison between the 3D and projected mass corrections has direct implications for the interpretation of XRISM observations. As shown in Sect.~\ref{sec:2d masses}, the non-thermal pressure fraction estimated from projections is systematically lower than its 3D counterpart by about a factor of two, and certain sightlines are strongly affected by the local filamentary environment. For a cluster like Virgo, XRISM measurements of the sightline velocity dispersion would thus only partially capture the non-thermal pressure support, and the resulting mass correction would be insufficient to fully account for the hydrostatic mass bias. In addition, the strong departure from isotropy quantified by $\beta$, see bottom panel of Fig.~\ref{vel_rad_prof_3D}, implies that the assumption $\sigma_{\rm 1D}^2=\sigma_{\rm 3D}^2/3$ used to recover the 3D velocity dispersion from line-of-sight measurements is itself a source of systematic error, which contributes to the underestimation of the non-thermal pressure fraction from projections.

Moreover, the deprojected hydrostatic mass is already significantly overestimated at $R_{500}$ due to the model-free geometrical deprojection method used in \citet{2024A&A...682A.157Lebeau}. Adding mass through either correction method further increases this overestimation, highlighting that correcting the hydrostatic mass for non-thermal pressure support is only meaningful if the hydrostatic mass itself is accurately estimated. The strong dependence of the results on the sightline further complicates the picture: the \textit{cen} sightline, aligned with the filaments connected to Virgo, yields a very different velocity dispersion and thus mass correction compared to the other sightlines. In practice, the correction of the hydrostatic mass from velocity measurements should thus be interpreted with particular caution for clusters whose sightline intersects prominent filaments of their large-scale environment, as is the case for Virgo.

It is also important to recall that this study is based on a single simulated cluster, and the results are specific to the dynamical state and local environment of the Virgo replica. Extending this analysis to a larger sample of clusters with different masses, formation histories, and environments would be necessary to assess the generality of our findings. Furthermore, the projected quantities used in this work are not mock observations: they do not include instrumental effects such as the point spread function, energy resolution, or photon noise, which could further degrade the accuracy of the velocity dispersion measurement. We plan to address both of these points in dedicated studies using mock XRISM observations and additional clusters from larger constrained simulations of the local Universe.

\section{Conclusion}
\label{conclusion}

In this work, we compare two approaches to correct the hydrostatic mass of galaxy clusters by accounting for non-thermal gas motions, using a constrained hydrodynamical zoom-in simulation of the Virgo cluster as a case study: the non-thermal pressure correction ($M_\alpha$) and the effective mass terms correction ($M_{\rm eff}$). We estimate these corrected masses both from 3D radial profiles in four distinct regions of the cluster and from projected sightline velocities mimicking XRISM-like measurements.

We show that the two correction methods are not equivalent: $M_\alpha$ increasingly exceeds $M_{\rm HE}$ with radius due to the rising non-thermal pressure fraction, whereas $M_{\rm eff}$ provides an almost radially constant correction due to the partial compensation between the velocity dispersion and bulk rotation terms. Both approaches increase the estimated mass by up to 25 to 40\% in the outskirts, but neither fully removes the hydrostatic mass bias. When estimated from projections, the two correction methods converge for a given sightline, but the non-thermal pressure fraction is underestimated by about a factor of two compared to the 3D case, and projection effects strongly affect the results, particularly the alignment of the sightline with cosmic filaments.

This case study does not aim to provide universal prescriptions for correcting the hydrostatic mass bias, but rather to demonstrate how the local environment, dynamical state, sightline orientation, and projection weighting schemes affect the mass correction in a detailed, well-characterized numerical replica. As XRISM already delivers increasingly precise measurements of the ICM velocity field, and as NewAthena \citep[][]{2025NatAs...9...36Cruise} will push these measurements further in the late 2030s, the correction methods investigated here will become directly applicable to observations. In future work, we plan to apply them to mock observations of this Virgo replica, including realistic instrumental effects, and to extend the analysis to other clusters in larger simulations of the local Universe.

\begin{acknowledgements}
The authors acknowledge the Gauss Centre for Supercomputing e.V. (www.gauss-centre.eu) for providing computing time on the GCS Supercomputers SuperMUC at LRZ Munich. This work was supported by the grant agreements ANR-21-CE31-0019 / 490702358 from the French Agence Nationale de la Recherche / DFG for the LOCALIZATION project. 
This work has been supported as part of France 2030 program ANR-11-IDEX-0003. 
S.E. acknowledges the financial contribution from {\it Theory Grant / Bando INAF per la Ricerca Fondamentale 2024} on "Constraining the non-thermal pressure in galaxy clusters with high-resolution X-ray spectroscopy" (1.05.24.05.10), and the support by the Jean D'Alembert fellowship program.
The authors thank Florent Renaud for sharing the {\tt rdramses} {\tt RAMSES} data reduction code. The authors thank François Roussel for his help in the verification of the {\tt rdramses} outputs. The Claude large language model from Anthropic was used in the final phase of this project to help refine the draft manuscript. The authors specify that all the ideas, analysis and writing were carried out by them and that they have taken the final decisions on the content of this article. 
   
\end{acknowledgements}

\bibliographystyle{aa} 
\bibliography{bibliography}

@ARTICLE{2025A&A...704A..14Lebeau,
       author = {{Lebeau}, Th{\'e}o and {Zaroubi}, Saleem and {Aghanim}, Nabila and {Sorce}, Jenny G. and {Langer}, Mathieu},
        title = "{Velocity fields and turbulence from cosmic filaments to galaxy clusters}",
      journal = {\aap},
         year = 2025,
        month = nov,
       volume = {704},
          eid = {A14},
        pages = {A14},
          doi = {10.1051/0004-6361/202553780},
archivePrefix = {arXiv},
       eprint = {2501.09573},
 primaryClass = {astro-ph.CO},
       adsurl = {https://ui.adsabs.harvard.edu/abs/2025A&amp;A...704A..14L}
}

@ARTICLE{2026A&A...707A.336Lebeau,
       author = {{Lebeau}, Th{\'e}o and {Ettori}, Stefano and {Sorce}, Jenny G. and {Aghanim}, Nabila and {Past{\'e}}, Jade},
        title = "{Gas motion in the intracluster medium of the Virgo cluster replica}",
      journal = {\aap},
         year = 2026,
        month = mar,
       volume = {707},
          eid = {A336},
        pages = {A336},
          doi = {10.1051/0004-6361/202555991},
archivePrefix = {arXiv},
       eprint = {2506.14441},
 primaryClass = {astro-ph.CO},
       adsurl = {https://ui.adsabs.harvard.edu/abs/2026A&A...707A.336L}
}

@article{gianfagna2021exploring,
  title={Exploring the hydrostatic mass bias in MUSIC clusters: application to the NIKA2 mock sample},
  author={Gianfagna, Giulia and De Petris, Marco and Yepes, Gustavo and De Luca, Federico and Sembolini, Federico and Cui, Weiguang and Biffi, Veronica and K{\'e}ruzor{\'e}, Florian and Mac{\'\i}as-P{\'e}rez, Juan and Mayet, Fr{\'e}d{\'e}ric and others},
  journal={\mnras},
  volume={502},
  number={4},
  pages={5115--5133},
  year={2021},
  publisher={Oxford University Press}
}

@article{planckcosmoparam2014,
  title={Planck 2013 results. XVI. Cosmological parameters},
  author={{Planck Collaboration}},
  journal={\aap},
  volume={571},
  pages={A16},
  year={2014}
}

@article{salvati2018constraints,
  title={Constraints from thermal Sunyaev-Zel’dovich cluster counts and power spectrum combined with CMB},
  author={Salvati, Laura and Douspis, Marian and Aghanim, Nabila},
  journal={\aap},
  volume={614},
  pages={A13},
  year={2018},
  publisher={EDP Sciences}
}

@article{dolag2008simulation,
  title={Simulation techniques for cosmological simulations},
  author={Dolag, Klaus and Borgani, Stefano and Schindler, Sabine and Diaferio, Antonio and Bykov, Andrei M},
  journal={Space science reviews},
  volume={134},
  number={1},
  pages={229--268},
  year={2008},
  publisher={Springer}
}

@article{sorce2021hydrodynamical,
  title={I--A hydrodynamical clone of the Virgo cluster of galaxies to confirm observationally driven formation scenarios},
  author={Sorce, Jenny G and Dubois, Yohan and Blaizot, J{\'e}r{\'e}my and McGee, Sean L and Yepes, Gustavo and Knebe, Alexander},
  journal={\mnras},
  volume={504},
  number={2},
  pages={2998--3012},
  year={2021},
  publisher={Oxford University Press}
}

@article{sorce2019virgo,
  title={Virgo: an unlikely cluster of galaxies because of its environment},
  author={Sorce, Jenny G and Blaizot, J{\'e}r{\'e}my and Dubois, Yohan},
  journal={\mnras},
  volume={486},
  number={3},
  pages={3951--3962},
  year={2019},
  publisher={Oxford University Press}
}

@article{teyssier2002cosmological,
  title={Cosmological hydrodynamics with adaptive mesh refinement-A new high resolution code called RAMSES},
  author={Teyssier, Romain},
  journal={\aap},
  volume={385},
  number={1},
  pages={337--364},
  year={2002},
  publisher={EDP Sciences}
}

@book{peebles2020large,
  title={The large-scale structure of the universe},
  author={Peebles, Phillip James Edwin},
  volume={98},
  year={2020},
  publisher={Princeton university press}
}

@article{sunyaev1972observations,
  title={The observations of relic radiation as a test of the nature of X-ray radiation from the clusters of galaxies},
  author={Sunyaev, RA and Zeldovich, Ya B},
  journal={Comments on Astrophysics and Space Physics},
  volume={4},
  pages={173},
  year={1972}
}

@article{dubois2016horizon,
  title={The HORIZON-AGN simulation: morphological diversity of galaxies promoted by AGN feedback},
  author={Dubois, Yohan and Peirani, S{\'e}bastien and Pichon, Christophe and Devriendt, Julien and Gavazzi, Rapha{\"e}l and Welker, Charlotte and Volonteri, Marta},
  journal={\mnras},
  volume={463},
  number={4},
  pages={3948--3964},
  year={2016},
  publisher={The Royal Astronomical Society}
}

@article{hoekstra2013masses,
  title={Masses of galaxy clusters from gravitational lensing},
  author={Hoekstra, Henk and Bartelmann, Matthias and Dahle, H{\aa}kon and Israel, Holger and Limousin, Marceau and Meneghetti, Massimo},
  journal={Space Science Reviews},
  volume={177},
  number={1},
  pages={75--118},
  year={2013},
  publisher={Springer}
}

@article{allen2011cosmological,
  title={Cosmological parameters from observations of galaxy clusters},
  author={Allen, Steven W and Evrard, August E and Mantz, Adam B},
  journal={Annual Review of Astronomy and Astrophysics},
  volume={49},
  pages={409--470},
  year={2011},
  publisher={Annual Reviews}
}

@article{vikhlinin2006chandra,
  title={Chandra sample of nearby relaxed galaxy clusters: Mass, gas fraction, and mass-temperature relation},
  author={Vikhlinin, Alexey and Kravtsov, Al and Forman, W and Jones, C and Markevitch, M and Murray, SS and Van Speybroeck, L},
  journal={\apj},
  volume={640},
  number={2},
  pages={691},
  year={2006},
  publisher={IOP Publishing}
}

@article{ettori2013mass,
  title={Mass profiles of Galaxy Clusters from X-ray analysis},
  author={Ettori, Stefano and Donnarumma, Annamaria and Pointecouteau, Etienne and Reiprich, Thomas H and Giodini, Stefania and Lovisari, Lorenzo and Schmidt, Robert W},
  journal={Space Science Reviews},
  volume={177},
  pages={119--154},
  year={2013},
  publisher={Springer}
}

@article{pratt2019galaxy,
  title={The galaxy cluster mass scale and its impact on cosmological constraints from the cluster population},
  author={Pratt, GW and Arnaud, M and Biviano, ANDREA and Eckert, D and Ettori, STEFANO and Nagai, D and Okabe, N and Reiprich, TH},
  journal={Space Science Reviews},
  volume={215},
  pages={1--82},
  year={2019},
  publisher={Springer}
}

@article{pearce2020hydrostatic,
  title={Hydrostatic mass estimates of massive galaxy clusters: a study with varying hydrodynamics flavours and non-thermal pressure support},
  author={Pearce, Francesca A and Kay, Scott T and Barnes, David J and Bower, Richard G and Schaller, Matthieu},
  journal={\mnras},
  volume={491},
  number={2},
  pages={1622--1642},
  year={2020},
  publisher={Oxford University Press}
}

@article{biffi2016nature,
  title={On the nature of hydrostatic equilibrium in galaxy clusters},
  author={Biffi, Ver{\'o}nica and Borgani, Stefano and Murante, Giuseppe and Rasia, Elena and Planelles, Susana and Granato, Gian Luigi and Ragone-Figueroa, Cinthia and Beck, Alexander M and Gaspari, Massimo and Dolag, Klaus},
  journal={\apj},
  volume={827},
  number={2},
  pages={112},
  year={2016},
  publisher={IOP Publishing}
}

@article{sorce2016cosmicflows,
  title={Cosmicflows constrained local universe simulations},
  author={Sorce, Jenny G and Gottl{\"o}ber, Stefan and Yepes, Gustavo and Hoffman, Yehuda and Courtois, Helene M and Steinmetz, Matthias and Tully, R Brent and Pomarede, Daniel and Carlesi, Edoardo},
  journal={\mnras},
  volume={455},
  number={2},
  pages={2078--2090},
  year={2016},
  publisher={Oxford University Press}
}

@article{nelson2014hydrodynamic,
  title={Hydrodynamic simulation of non-thermal pressure profiles of galaxy clusters},
  author={Nelson, Kaylea and Lau, Erwin T and Nagai, Daisuke},
  journal={\apj},
  volume={792},
  number={1},
  pages={25},
  year={2014},
  publisher={IOP Publishing}
}

@article{dubois2014dancing,
  title={Dancing in the dark: galactic properties trace spin swings along the cosmic web},
  author={Dubois, Yohan and Pichon, Christophe and Welker, Charlotte and Le Borgne, Damien and Devriendt, Julien and Laigle, Clotilde and Codis, Sandrine and Pogosyan, Dmitry and Arnouts, St{\'e}phane and Benabed, Karim and others},
  journal={\mnras},
  volume={444},
  number={2},
  pages={1453--1468},
  year={2014},
  publisher={Oxford University Press}
}

@article{dubois2021introducing,
  title={Introducing the NEWHORIZON simulation: Galaxy properties with resolved internal dynamics across cosmic time},
  author={Dubois, Yohan and Beckmann, Ricarda and Bournaud, Fr{\'e}d{\'e}ric and Choi, Hoseung and Devriendt, Julien and Jackson, Ryan and Kaviraj, Sugata and Kimm, Taysun and Kraljic, Katarina and Laigle, Clotilde and others},
  journal={\aap},
  volume={651},
  pages={A109},
  year={2021},
  publisher={EDP Sciences}
}

@article{rasia2004dynamical,
  title={A dynamical model for the distribution of dark matter and gas in galaxy clusters},
  author={Rasia, Elena and Tormen, Giuseppe and Moscardini, Lauro},
  journal={\mnras},
  volume={351},
  number={1},
  pages={237--252},
  year={2004},
  publisher={The Royal Astronomical Society}
}

@article{rasia2006systematics,
  title={Systematics in the X-ray cluster mass estimators},
  author={Rasia, Elena and Ettori, Stefano and Moscardini, L and Mazzotta, P and Borgani, Stefano and Dolag, K and Tormen, G and Cheng, LM and Diaferio, Antonaldo},
  journal={\mnras},
  volume={369},
  number={4},
  pages={2013--2024},
  year={2006},
  publisher={Blackwell Publishing Ltd Oxford, UK}
}

@article{kaiser1986evolution,
  title={Evolution and clustering of rich clusters},
  author={Kaiser, Nick},
  journal={\mnras},
  volume={222},
  number={2},
  pages={323--345},
  year={1986},
  publisher={The Royal Astronomical Society}
}

@article{shi2016locations,
  title={Locations of accretion shocks around galaxy clusters and the ICM properties: insights from self-similar spherical collapse with arbitrary mass accretion rates},
  author={Shi, Xun},
  journal={\mnras},
  volume={461},
  number={2},
  pages={1804--1815},
  year={2016},
  publisher={Oxford University Press}
}

@article{tully2013cosmicflows,
  title={Cosmicflows-2: the data},
  author={Tully, R Brent and Courtois, H{\'e}l{\`e}ne M and Dolphin, Andrew E and Fisher, J Richard and H{\'e}raudeau, Philippe and Jacobs, Bradley A and Karachentsev, Igor D and Makarov, Dmitry and Makarova, Lidia and Mitronova, Sofia and others},
  journal={The Astronomical Journal},
  volume={146},
  number={4},
  pages={86},
  year={2013},
  publisher={IOP Publishing}
}

@article{gaspari2013constraining,
  title={Constraining turbulence and conduction in the hot ICM through density perturbations},
  author={Gaspari, M and Churazov, E},
  journal={\aap},
  volume={559},
  pages={A78},
  year={2013},
  publisher={EDP Sciences}
}

@article{valles2021troubled,
  title={Troubled cosmic flows: turbulence, enstrophy, and helicity from the assembly history of the intracluster medium},
  author={Vall{\'e}s-P{\'e}rez, David and Planelles, Susana and Quilis, Vicent},
  journal={\mnras},
  volume={504},
  number={1},
  pages={510--527},
  year={2021},
  publisher={Oxford University Press}
}

@article{schuecker2004probing,
  title={Probing turbulence in the Coma galaxy cluster},
  author={Schuecker, Peter and Finoguenov, A and Miniati, F and B{\"o}hringer, H and Briel, UG},
  journal={\aap},
  volume={426},
  number={2},
  pages={387--397},
  year={2004},
  publisher={EDP Sciences}
}

@article{eckert2019non,
  title={Non-thermal pressure support in X-COP galaxy clusters},
  author={Eckert, D and Ghirardini, V and Ettori, Stefano and Rasia, Elena and Biffi, V and Pointecouteau, E and Rossetti, Mariachiara and Molendi, Silvano and Vazza, Franco and Gastaldello, Fabio and others},
  journal={\aap},
  volume={621},
  pages={A40},
  year={2019},
  publisher={EDP Sciences}
}

@ARTICLE{2023A&A...673A..91Dupourque,
       author = {{Dupourqu{\'e}}, S. and {Clerc}, N. and {Pointecouteau}, E. and {Eckert}, D. and {Ettori}, S. and {Vazza}, F.},
        title = "{Investigating the turbulent hot gas in X-COP galaxy clusters}",
      journal = {\aap},
         year = 2023,
        month = may,
       volume = {673},
          eid = {A91},
        pages = {A91},
          doi = {10.1051/0004-6361/202245779},
archivePrefix = {arXiv},
       eprint = {2303.15102},
 primaryClass = {astro-ph.CO},
       adsurl = {https://ui.adsabs.harvard.edu/abs/2023A&A...673A..91D}
}

@article{dupourque2024chex,
  title={CHEX-MATE: Turbulence in the intra-cluster medium from X-ray surface brightness fluctuations},
  author={Dupourqu{\'e}, S and Clerc, N and Pointecouteau, E and Eckert, D and Gaspari, M and Lovisari, L and Pratt, GW and Rasia, E and Rossetti, M and Vazza, F and others},
  journal={\aap},
  volume={687},
  pages={A58},
  year={2024},
  publisher={EDP Sciences}
}

@ARTICLE{romero2023inf,
       author = {{Romero}, Charles E. and {Gaspari}, Massimo and {Schellenberger}, Gerrit and {Bhandarkar}, Tanay and {Devlin}, Mark and {Dicker}, Simon R. and {Forman}, William and {Khatri}, Rishi and {Kraft}, Ralph and {Di Mascolo}, Luca and {Mason}, Brian S. and {Moravec}, Emily and {Mroczkowski}, Tony and {Nulsen}, Paul and {Orlowski-Scherer}, John and {Perez Sarmiento}, Karen and {Sarazin}, Craig and {Sievers}, Jonathan and {Su}, Yuanyuan},
        title = "{Inferences from Surface Brightness Fluctuations of Zwicky 3146 via the Sunyaev-Zel'dovich Effect and X-Ray Observations}",
      journal = {\apj},
         year = 2023,
        month = jul,
       volume = {951},
       number = {1},
          eid = {41},
        pages = {41},
          doi = {10.3847/1538-4357/acd3f0},
archivePrefix = {arXiv},
       eprint = {2305.05790},
 primaryClass = {astro-ph.CO},
       adsurl = {https://ui.adsabs.harvard.edu/abs/2023ApJ...951...41R}
}

@ARTICLE{romero2024surface,
       author = {{Romero}, Charles E. and {Gaspari}, Massimo and {Schellenberger}, Gerrit and {Benson}, Bradford A. and {Bleem}, Lindsey E. and {Bulbul}, Esra and {Klein}, Matthias and {Kraft}, Ralph and {Nulsen}, Paul and {Reichardt}, Christian L. and {Salvati}, Laura and {Somboonpanyakul}, Taweewat and {Su}, Yuanyuan},
        title = "{Surface Brightness Fluctuations in Two SPT Clusters: A Pilot Study}",
      journal = {\apj},
         year = 2024,
        month = jul,
       volume = {970},
       number = {1},
          eid = {73},
        pages = {73},
          doi = {10.3847/1538-4357/ad2992},
archivePrefix = {arXiv},
       eprint = {2404.04373},
 primaryClass = {astro-ph.CO},
       adsurl = {https://ui.adsabs.harvard.edu/abs/2024ApJ...970...73R}
}

@ARTICLE{roncarelli2018athena,
       author = {{Roncarelli}, M. and {Gaspari}, M. and {Ettori}, S. and {Biffi}, V. and {Brighenti}, F. and {Bulbul}, E. and {Clerc}, N. and {Cucchetti}, E. and {Pointecouteau}, E. and {Rasia}, E.},
        title = "{Measuring turbulence and gas motions in galaxy clusters via synthetic Athena X-IFU observations}",
      journal = {\aap},
         year = 2018,
        month = oct,
       volume = {618},
          eid = {A39},
        pages = {A39},
          doi = {10.1051/0004-6361/201833371},
archivePrefix = {arXiv},
       eprint = {1805.02577},
 primaryClass = {astro-ph.CO},
       adsurl = {https://ui.adsabs.harvard.edu/abs/2018A&A...618A..39R}
}

@ARTICLE{2009ApJ...705.1129Lau,
       author = {{Lau}, Erwin T. and {Kravtsov}, Andrey V. and {Nagai}, Daisuke},
        title = "{Residual Gas Motions in the Intracluster Medium and Bias in Hydrostatic Measurements of Mass Profiles of Clusters}",
      journal = {\apj},
         year = 2009,
        month = nov,
       volume = {705},
       number = {2},
        pages = {1129-1138},
          doi = {10.1088/0004-637X/705/2/1129},
archivePrefix = {arXiv},
       eprint = {0903.4895},
 primaryClass = {astro-ph.CO},
       adsurl = {https://ui.adsabs.harvard.edu/abs/2009ApJ...705.1129L}
}

@ARTICLE{2013ApJ...777..151Lau,
       author = {{Lau}, Erwin T. and {Nagai}, Daisuke and {Nelson}, Kaylea},
        title = "{Weighing Galaxy Clusters with Gas. I. On the Methods of Computing Hydrostatic Mass Bias}",
      journal = {\apj},
         year = 2013,
        month = nov,
       volume = {777},
       number = {2},
          eid = {151},
        pages = {151},
          doi = {10.1088/0004-637X/777/2/151},
archivePrefix = {arXiv},
       eprint = {1306.3993},
 primaryClass = {astro-ph.CO},
       adsurl = {https://ui.adsabs.harvard.edu/abs/2013ApJ...777..151L}
}

@ARTICLE{2018PASJ...70...51Ota,
       author = {{Ota}, Naomi and {Nagai}, Daisuke and {Lau}, Erwin T.},
        title = "{Constraining hydrostatic mass bias of galaxy clusters with high-resolution X-ray spectroscopy}",
      journal = {\pasj},
         year = 2018,
        month = jun,
       volume = {70},
       number = {3},
          eid = {51},
        pages = {51},
          doi = {10.1093/pasj/psy040},
archivePrefix = {arXiv},
       eprint = {1507.02730},
 primaryClass = {astro-ph.CO},
       adsurl = {https://ui.adsabs.harvard.edu/abs/2018PASJ...70...51O}
}

@ARTICLE{2013ApJ...767...79Suto,
       author = {{Suto}, Daichi and {Kawahara}, Hajime and {Kitayama}, Tetsu and {Sasaki}, Shin and {Suto}, Yasushi and {Cen}, Renyue},
        title = "{Validity of Hydrostatic Equilibrium in Galaxy Clusters from Cosmological Hydrodynamical Simulations}",
      journal = {\apj},
         year = 2013,
        month = apr,
       volume = {767},
       number = {1},
          eid = {79},
        pages = {79},
          doi = {10.1088/0004-637X/767/1/79},
archivePrefix = {arXiv},
       eprint = {1302.5172},
 primaryClass = {astro-ph.CO},
       adsurl = {https://ui.adsabs.harvard.edu/abs/2013ApJ...767...79S}
}

@misc{binney2008galactic,
  title={Galactic Dynamics},
  author={Binney, James and Tremaine, Scott},
  year={2008},
  publisher={Princeton University Press}
}

@ARTICLE{2019SSRv..215...24Simionescu,
       author = {{Simionescu}, Aurora and {ZuHone}, John and {Zhuravleva}, Irina and {Churazov}, Eugene and {Gaspari}, Massimo and {Nagai}, Daisuke and {Werner}, Norbert and {Roediger}, Elke and {Canning}, Rebecca and {Eckert}, Dominique and {Gu}, Liyi and {Paerels}, Frits},
        title = "{Constraining Gas Motions in the Intra-Cluster Medium}",
      journal = {\ssr},
         year = 2019,
        month = feb,
       volume = {215},
       number = {2},
          eid = {24},
        pages = {24},
          doi = {10.1007/s11214-019-0590-1},
archivePrefix = {arXiv},
       eprint = {1902.00024},
 primaryClass = {astro-ph.CO},
       adsurl = {https://ui.adsabs.harvard.edu/abs/2019SSRv..215...24S}
}

@ARTICLE{2024A&A...690A.238Aymerich,
       author = {{Aymerich}, G. and {Douspis}, M. and {Pratt}, G.~W. and {Salvati}, L. and {Soubri{\'e}}, E. and {Andrade-Santos}, F. and {Forman}, W.~R. and {Jones}, C. and {Aghanim}, N. and {Kraft}, R. and {van Weeren}, R.~J.},
        title = "{Cosmological constraints from the Planck cluster catalogue with new multi-wavelength mass calibration from Chandra and CFHT}",
      journal = {\aap},
         year = 2024,
        month = oct,
       volume = {690},
          eid = {A238},
        pages = {A238},
          doi = {10.1051/0004-6361/202449513},
archivePrefix = {arXiv},
       eprint = {2402.04006},
 primaryClass = {astro-ph.CO},
       adsurl = {https://ui.adsabs.harvard.edu/abs/2024A&A...690A.238A}
}

@ARTICLE{2025A&A...693A.263Groth,
       author = {{Groth}, Frederick and {Valentini}, Milena and {Steinwandel}, Ulrich P. and {Vall{\'e}s-P{\'e}rez}, David and {Dolag}, Klaus},
        title = "{Turbulent pressure support in galaxy clusters: Impact of the hydrodynamical solver}",
      journal = {\aap},
         year = 2025,
        month = jan,
       volume = {693},
          eid = {A263},
        pages = {A263},
          doi = {10.1051/0004-6361/202451803},
archivePrefix = {arXiv},
       eprint = {2408.02719},
 primaryClass = {astro-ph.CO},
       adsurl = {https://ui.adsabs.harvard.edu/abs/2025A&A...693A.263G}
}

@ARTICLE{2025ApJ...982L...5XRISM,
       author = {{XRISM Collaboration} and {Audard}, Marc and {Awaki}, Hisamitsu and {Ballhausen}, Ralf and {Bamba}, Aya and {Behar}, Ehud and {Boissay-Malaquin}, Rozenn and {Brenneman}, Laura and {Brown}, Gregory V. and {Corrales}, Lia and {Costantini}, Elisa and {Cumbee}, Renata and {Diaz Trigo}, Maria and {Done}, Chris and {Dotani}, Tadayasu and {Ebisawa}, Ken and {Eckart}, Megan E. and {Eckert}, Dominique and {Eguchi}, Satoshi and {Enoto}, Teruaki and {Ezoe}, Yuichiro and {Foster}, Adam and {Fujimoto}, Ryuichi and {Fujita}, Yutaka and {Fukazawa}, Yasushi and {Fukushima}, Kotaro and {Furuzawa}, Akihiro and {Gallo}, Luigi and {Garc{\'\i}a}, Javier A. and {Gu}, Liyi and {Guainazzi}, Matteo and {Hagino}, Kouichi and {Hamaguchi}, Kenji and {Hatsukade}, Isamu and {Hayashi}, Katsuhiro and {Hayashi}, Takayuki and {Hell}, Natalie and {Hodges-Kluck}, Edmund and {Hornschemeier}, Ann and {Ichinohe}, Yuto and {Ishida}, Manabu and {Ishikawa}, Kumi and {Ishisaki}, Yoshitaka and {Kaastra}, Jelle and {Kallman}, Timothy and {Kara}, Erin and {Katsuda}, Satoru and {Kanemaru}, Yoshiaki and {Kelley}, Richard and {Kilbourne}, Caroline and {Kitamoto}, Shunji and {Kobayashi}, Shogo and {Kohmura}, Takayoshi and {Kubota}, Aya and {Leutenegger}, Maurice and {Loewenstein}, Michael and {Maeda}, Yoshitomo and {Markevitch}, Maxim and {Matsumoto}, Hironori and {Matsushita}, Kyoko and {McCammon}, Dan and {McNamara}, Brian and {Mernier}, Fran{\c{c}}ois and {Miller}, Eric D. and {Miller}, Jon M. and {Mitsuishi}, Ikuyuki and {Mizumoto}, Misaki and {Mizuno}, Tsunefumi and {Mori}, Koji and {Mukai}, Koji and {Murakami}, Hiroshi and {Mushotzky}, Richard and {Nakajima}, Hiroshi and {Nakazawa}, Kazuhiro and {Ness}, Jan-Uwe and {Nobukawa}, Kumiko and {Nobukawa}, Masayoshi and {Noda}, Hirofumi and {Odaka}, Hirokazu and {Ogawa}, Shoji and {Ogorzalek}, Anna and {Okajima}, Takashi and {Ota}, Naomi and {Paltani}, Stephane and {Petre}, Robert and {Plucinsky}, Paul and {Porter}, Frederick S. and {Pottschmidt}, Katja and {Sato}, Kosuke and {Sato}, Toshiki and {Sawada}, Makoto and {Seta}, Hiromi and {Shidatsu}, Megumi and {Simionescu}, Aurora and {Smith}, Randall and {Suzuki}, Hiromasa and {Szymkowiak}, Andrew and {Takahashi}, Hiromitsu and {Takeo}, Mai and {Tamagawa}, Toru and {Tamura}, Keisuke and {Tanaka}, Takaaki and {Tanimoto}, Atsushi and {Tashiro}, Makoto and {Terada}, Yukikatsu and {Terashima}, Yuichi and {Tsuboi}, Yohko and {Tsujimoto}, Masahiro and {Tsunemi}, Hiroshi and {Tsuru}, Takeshi and {Uchida}, Hiroyuki and {Uchida}, Nagomi and {Uchida}, Yuusuke and {Uchiyama}, Hideki and {Ueda}, Yoshihiro and {Uno}, Shinichiro and {Vink}, Jacco and {Watanabe}, Shin and {Williams}, Brian J. and {Yamada}, Satoshi and {Yamada}, Shinya and {Yamaguchi}, Hiroya and {Yamaoka}, Kazutaka and {Yamasaki}, Noriko and {Yamauchi}, Makoto and {Yamauchi}, Shigeo and {Yaqoob}, Tahir and {Yoneyama}, Tomokage and {Yoshida}, Tessei and {Yukita}, Mihoko and {Zhuravleva}, Irina and {Bartalesi}, Tommaso and {Ettori}, Stefano and {Kosarzycki}, Roman and {Lovisari}, Lorenzo and {Rose}, Tom and {Sarkar}, Arnab and {Sun}, Ming and {Tamhane}, Prathamesh},
        title = "{XRISM Reveals Low Nonthermal Pressure in the Core of the Hot, Relaxed Galaxy Cluster A2029}",
      journal = {ApJL},
         year = 2025,
        month = mar,
       volume = {982},
       number = {1},
          eid = {L5},
        pages = {L5},
          doi = {10.3847/2041-8213/ada7cd},
archivePrefix = {arXiv},
       eprint = {2501.05514},
 primaryClass = {astro-ph.HE},
       adsurl = {https://ui.adsabs.harvard.edu/abs/2025ApJ...982L...5X}
}

@book{liddle2015introduction,
  title={An introduction to modern cosmology},
  author={Liddle, Andrew},
  year={2015},
  publisher={John Wiley \& Sons}
}

@ARTICLE{2013SSRv..177..247Giodini,
       author = {{Giodini}, S. and {Lovisari}, L. and {Pointecouteau}, E. and {Ettori}, S. and {Reiprich}, T.~H. and {Hoekstra}, H.},
        title = "{Scaling Relations for Galaxy Clusters: Properties and Evolution}",
      journal = {SSR},
         year = 2013,
        month = aug,
       volume = {177},
       number = {1-4},
        pages = {247-282},
          doi = {10.1007/s11214-013-9994-5},
archivePrefix = {arXiv},
       eprint = {1305.3286},
 primaryClass = {astro-ph.CO},
       adsurl = {https://ui.adsabs.harvard.edu/abs/2013SSRv..177..247G}
}

@ARTICLE{2007ApJ...668....1Nagai,
       author = {{Nagai}, Daisuke and {Kravtsov}, Andrey V. and {Vikhlinin}, Alexey},
        title = "{Effects of Galaxy Formation on Thermodynamics of the Intracluster Medium}",
      journal = {ApJ},
         year = 2007,
        month = oct,
       volume = {668},
       number = {1},
        pages = {1-14},
          doi = {10.1086/521328},
archivePrefix = {arXiv},
       eprint = {astro-ph/0703661},
 primaryClass = {astro-ph},
       adsurl = {https://ui.adsabs.harvard.edu/abs/2007ApJ...668....1N}
}

@ARTICLE{2005JCAP...01..009DolagUHECRs,
       author = {{Dolag}, Klaus and {Grasso}, Dario and {Springel}, Volker and {Tkachev}, Igor},
        title = "{Constrained simulations of the magnetic field in the local Universe and the propagation of ultrahigh energy cosmic rays}",
      journal = {JCAP},
         year = 2005,
        month = jan,
       volume = {2005},
       number = {1},
          eid = {009},
        pages = {009},
          doi = {10.1088/1475-7516/2005/01/009},
archivePrefix = {arXiv},
       eprint = {astro-ph/0410419},
 primaryClass = {astro-ph},
       adsurl = {https://ui.adsabs.harvard.edu/abs/2005JCAP...01..009D}
}

@INPROCEEDINGS{2012IJMPS..12..280Ferrari,
       author = {{Ferrari}, Chiara},
        title = "{Non-Thermal Phenomena in Galaxy Clusters}",
    booktitle = {Int. J. Mod. Phys.: Conf. Ser.},
         year = 2012,
       series = {Int. J. Mod. Phys.: Conf. Ser.},
       volume = {12},
        month = mar,
        pages = {280-289},
          doi = {10.1142/S2010194512006484},
       adsurl = {https://ui.adsabs.harvard.edu/abs/2012IJMPS..12..280F}
}

@ARTICLE{2019SSRv..215...16VanWeeren,
       author = {{Van Weeren}, R.~J. and {de Gasperin}, F. and {Akamatsu}, H. and {Br{\"u}ggen}, M. and {Feretti}, L. and {Kang}, H. and {Stroe}, A. and {Zandanel}, F.},
        title = "{Diffuse Radio Emission from Galaxy Clusters}",
      journal = {SSR},
         year = 2019,
        month = feb,
       volume = {215},
       number = {1},
          eid = {16},
        pages = {16},
          doi = {10.1007/s11214-019-0584-z},
archivePrefix = {arXiv},
       eprint = {1901.04496},
 primaryClass = {astro-ph.HE},
       adsurl = {https://ui.adsabs.harvard.edu/abs/2019SSRv..215...16V}
}

@ARTICLE{2024A&A...682A..45Lovisari,
       author = {{Lovisari}, L. and {Ettori}, S. and {Rasia}, E. and {Gaspari}, M. and {Bourdin}, H. and {Campitiello}, M.~G. and {Rossetti}, M. and {Bartalucci}, I. and {De Grandi}, S. and {De Luca}, F. and {De Petris}, M. and {Eckert}, D. and {Forman}, W. and {Gastaldello}, F. and {Ghizzardi}, S. and {Jones}, C. and {Kay}, S. and {Kim}, J. and {Maughan}, B.~J. and {Mazzotta}, P. and {Pointecouteau}, E. and {Pratt}, G.~W. and {Sayers}, J. and {Sereno}, M. and {Simonte}, M. and {Tozzi}, P.},
        title = "{CHEX-MATE: Characterization of the intra-cluster medium temperature distribution}",
      journal = {\aap},
         year = 2024,
        month = feb,
       volume = {682},
          eid = {A45},
        pages = {A45},
          doi = {10.1051/0004-6361/202346651},
archivePrefix = {arXiv},
       eprint = {2311.02176},
 primaryClass = {astro-ph.CO},
       adsurl = {https://ui.adsabs.harvard.edu/abs/2024A&A...682A..45L}
}

@ARTICLE{2025PASJ...77S.242XRISMA2029,
       author = {{XRISM Collaboration} and {Audard}, Marc and {Awaki}, Hisamitsu and {Ballhausen}, Ralf and {Bamba}, Aya and {Behar}, Ehud and {Boissay-Malaquin}, Rozenn and {Brenneman}, Laura and {Brown}, Gregory V. and {Corrales}, Lia and {Costantini}, Elisa and {Cumbee}, Renata and {D{\'\i}az Trigo}, Maria and {Done}, Chris and {Dotani}, Tadayasu and {Ebisawa}, Ken and {Eckart}, Megan E. and {Eckert}, Dominique and {Eguchi}, Satoshi and {Enoto}, Teruaki and {Ezoe}, Yuichiro and {Foster}, Adam and {Fujimoto}, Ryuichi and {Fujita}, Yutaka and {Fukazawa}, Yasushi and {Fukushima}, Kotaro and {Furuzawa}, Akihiro and {Gallo}, Luigi C. and {Garc{\'\i}a}, Javier A. and {Gu}, Liyi and {Guainazzi}, Matteo and {Hagino}, Kouichi and {Hamaguchi}, Kenji and {Hatsukade}, Isamu and {Hayashi}, Katsuhiro and {Hayashi}, Takayuki and {Hell}, Natalie and {Hodges-Kluck}, Edmund and {Hornschemeier}, Ann and {Ichinohe}, Yuto and {Ishi}, Daiki and {Ishida}, Manabu and {Ishikawa}, Kumi and {Ishisaki}, Yoshitaka and {Kaastra}, Jelle and {Kallman}, Timothy and {Kara}, Erin and {Katsuda}, Satoru and {Kanemaru}, Yoshiaki and {Kelley}, Richard and {Kilbourne}, Caroline and {Kitamoto}, Shunji and {Kobayashi}, Shogo and {Kohmura}, Takayoshi and {Kubota}, Aya and {Leutenegger}, Maurice A. and {Loewenstein}, Michael and {Maeda}, Yoshitomo and {Markevitch}, Maxim and {Matsumoto}, Hironori and {Matsushita}, Kyoko and {McCammon}, Dan and {McNamara}, Brian and {Mernier}, Fran{\c{c}}ois and {Miller}, Eric and {Miller}, Jon M. and {Mitsuishi}, Ikuyuki and {Mizumoto}, Misaki and {Mizuno}, Tsunefumi and {Mori}, Koji and {Mukai}, Koji and {Murakami}, Hiroshi and {Mushotzky}, Richard and {Nakajima}, Hiroshi and {Nakazawa}, Kazuhiro and {Ness}, Jan-Uwe and {Nobukawa}, Kumiko and {Nobukawa}, Masayoshi and {Noda}, Hirofumi and {Odaka}, Hirokazu and {Ogawa}, Shoji and {Ogorzalek}, Anna and {Okajima}, Takashi and {Ota}, Naomi and {Paltani}, Stephane and {Petre}, Robert and {Plucinsky}, Paul and {Porter}, Frederick S. and {Pottschmidt}, Katja and {Sato}, Kosuke and {Sato}, Toshiki and {Sawada}, Makoto and {Seta}, Hiromi and {Shidatsu}, Megumi and {Simionescu}, Aurora and {Smith}, Randall and {Suzuki}, Hiromasa and {Szymkowiak}, Andrew and {Takahashi}, Hiromitsu and {Takeo}, Mai and {Tamagawa}, Toru and {Tamura}, Keisuke and {Tanaka}, Takaaki and {Tanimoto}, Atsushi and {Tashiro}, Makoto and {Terada}, Yukikatsu and {Terashima}, Yuichi and {Tsuboi}, Yohko and {Tsujimoto}, Masahiro and {Tsunemi}, Hiroshi and {Tsuru}, Takeshi Go and {Uchida}, Hiroyuki and {Uchida}, Nagomi and {Uchida}, Yuusuke and {Uchiyama}, Hideki and {Ueda}, Yoshihiro and {Uno}, Shinichiro and {Vink}, Jacco and {Watanabe}, Shin and {Williams}, Brian J. and {Yamada}, Satoshi and {Yamada}, Shinya and {Yamaguchi}, Hiroya and {Yamaoka}, Kazutaka and {Yamasaki}, Noriko and {Yamauchi}, Makoto and {Yamauchi}, Shigeo and {Yaqoob}, Tahir and {Yoneyama}, Tomokage and {Yoshida}, Tessei and {Yukita}, Mihoko and {Zhuravleva}, Irina and {Bartalesi}, Tommaso and {Ettori}, Stefano and {Kosarzycki}, Roman and {Lovisari}, Lorenzo and {Rose}, Tom and {Sarkar}, Arnab and {Sun}, Ming and {Tamhane}, Prathamesh},
        title = "{Constraining gas motion and non-thermal pressure beyond the core of the Abell 2029 galaxy cluster with XRISM}",
      journal = {\pasj},
         year = 2025,
        month = sep,
       volume = {77},
        pages = {S242-S253},
          doi = {10.1093/pasj/psaf055},
archivePrefix = {arXiv},
       eprint = {2505.06533},
 primaryClass = {astro-ph.CO},
       adsurl = {https://ui.adsabs.harvard.edu/abs/2025PASJ...77S.242X}
}

@ARTICLE{2025ApJ...985L..20XRISM-Coma,
       author = {{XRISM Collaboration} and {Audard}, Marc and {Awaki}, Hisamitsu and {Ballhausen}, Ralf and {Bamba}, Aya and {Behar}, Ehud and {Boissay-Malaquin}, Rozenn and {Brenneman}, Laura and {Brown}, Gregory V. and {Corrales}, Lia and {Costantini}, Elisa and {Cumbee}, Renata and {Diaz Trigo}, Maria and {Done}, Chris and {Dotani}, Tadayasu and {Ebisawa}, Ken and {Eckart}, Megan E. and {Eckert}, Dominique and {Eguchi}, Satoshi and {Enoto}, Teruaki and {Ezoe}, Yuichiro and {Foster}, Adam and {Fujimoto}, Ryuichi and {Fujita}, Yutaka and {Fukazawa}, Yasushi and {Fukushima}, Kotaro and {Furuzawa}, Akihiro and {Gallo}, Luigi and {Garc{\'\i}a}, Javier A. and {Gu}, Liyi and {Guainazzi}, Matteo and {Hagino}, Kouichi and {Hamaguchi}, Kenji and {Hatsukade}, Isamu and {Hayashi}, Katsuhiro and {Hayashi}, Takayuki and {Hell}, Natalie and {Hodges-Kluck}, Edmund and {Hornschemeier}, Ann and {Ichinohe}, Yuto and {Ishi}, Daiki and {Ishida}, Manabu and {Ishikawa}, Kumi and {Ishisaki}, Yoshitaka and {Kaastra}, Jelle and {Kallman}, Timothy and {Kara}, Erin and {Katsuda}, Satoru and {Kanemaru}, Yoshiaki and {Kelley}, Richard and {Kilbourne}, Caroline and {Kitamoto}, Shunji and {Kobayashi}, Shogo and {Kohmura}, Takayoshi and {Kubota}, Aya and {Leutenegger}, Maurice and {Loewenstein}, Michael and {Maeda}, Yoshitomo and {Markevitch}, Maxim and {Matsumoto}, Hironori and {Matsushita}, Kyoko and {McCammon}, Dan and {McNamara}, Brian and {Mernier}, Fran{\c{c}}ois and {Miller}, Eric D. and {Miller}, Jon M. and {Mitsuishi}, Ikuyuki and {Mizumoto}, Misaki and {Mizuno}, Tsunefumi and {Mori}, Koji and {Mukai}, Koji and {Murakami}, Hiroshi and {Mushotzky}, Richard and {Nakajima}, Hiroshi and {Nakazawa}, Kazuhiro and {Ness}, Jan-Uwe and {Nobukawa}, Kumiko and {Nobukawa}, Masayoshi and {Noda}, Hirofumi and {Odaka}, Hirokazu and {Ogawa}, Shoji and {Ogorza{\l}ek}, Anna and {Okajima}, Takashi and {Ota}, Naomi and {Paltani}, St{\'e}phane and {Petre}, Robert and {Plucinsky}, Paul and {Porter}, Frederick S. and {Pottschmidt}, Katja and {Sato}, Kosuke and {Sato}, Toshiki and {Sawada}, Makoto and {Seta}, Hiromi and {Shidatsu}, Megumi and {Simionescu}, Aurora and {Smith}, Randall and {Suzuki}, Hiromasa and {Szymkowiak}, Andrew and {Takahashi}, Hiromitsu and {Takeo}, Mai and {Tamagawa}, Toru and {Tamura}, Keisuke and {Tanaka}, Takaaki and {Tanimoto}, Atsushi and {Tashiro}, Makoto and {Terada}, Yukikatsu and {Terashima}, Yuichi and {Tsuboi}, Yohko and {Tsujimoto}, Masahiro and {Tsunemi}, Hiroshi and {Tsuru}, Takeshi and {T{\"u}mer}, Ay{\c{s}}eg{\"u}l and {Uchida}, Hiroyuki and {Uchida}, Nagomi and {Uchida}, Yuusuke and {Uchiyama}, Hideki and {Ueda}, Shutaro and {Ueda}, Yoshihiro and {Uno}, Shinichiro and {Vink}, Jacco and {Watanabe}, Shin and {Williams}, Brian J. and {Yamada}, Satoshi and {Yamada}, Shinya and {Yamaguchi}, Hiroya and {Yamaoka}, Kazutaka and {Yamasaki}, Noriko and {Yamauchi}, Makoto and {Yamauchi}, Shigeo and {Yaqoob}, Tahir and {Yoneyama}, Tomokage and {Yoshida}, Tessei and {Yukita}, Mihoko and {Zhuravleva}, Irina and {Fabian}, Andrew and {Nelson}, Dylan and {Okabe}, Nobuhiro and {Pillepich}, Annalisa and {Potter}, Cicely and {Regamey}, Manon and {Sakai}, Kosei and {Shishido}, Mona and {Truong}, Nhut and {Wik}, Daniel R. and {Zuhone}, John},
        title = "{XRISM Forecast for the Coma Cluster: Stormy, with a Steep Power Spectrum}",
      journal = {ApJL},
         year = 2025,
        month = may,
       volume = {985},
       number = {1},
          eid = {L20},
        pages = {L20},
          doi = {10.3847/2041-8213/add2f6},
archivePrefix = {arXiv},
       eprint = {2504.20928},
 primaryClass = {astro-ph.HE},
       adsurl = {https://ui.adsabs.harvard.edu/abs/2025ApJ...985L..20X}
}

@ARTICLE{2018MNRAS.481L.120Vazza,
       author = {{Vazza}, F. and {Angelinelli}, M. and {Jones}, T.~W. and {Eckert}, D. and {Br{\"u}ggen}, M. and {Brunetti}, G. and {Gheller}, C.},
        title = "{The turbulent pressure support in galaxy clusters revisited}",
      journal = {MNRAS},
         year = 2018,
        month = nov,
       volume = {481},
       number = {1},
        pages = {L120-L124},
          doi = {10.1093/mnrasl/sly172},
archivePrefix = {arXiv},
       eprint = {1809.02690},
 primaryClass = {astro-ph.CO},
       adsurl = {https://ui.adsabs.harvard.edu/abs/2018MNRAS.481L.120V}
}

@article{mamon2013mamposst,
  title={MAMPOSSt: Modelling anisotropy and mass profiles of observed spherical systems--I. Gaussian 3D velocities},
  author={Mamon, Gary A and Biviano, Andrea and Bou{\'e}, Gwena{\"e}l},
  journal={MNRAS},
  volume={429},
  number={4},
  pages={3079--3098},
  year={2013},
  publisher={Oxford University Press}
}

@article{diaferio1997infall,
  title={Infall regions of galaxy clusters},
  author={Diaferio, Antonaldo and Geller, Margaret J},
  journal={ApJ},
  volume={481},
  number={2},
  pages={633},
  year={1997},
  publisher={IOP Publishing}
}

@article{okabe2016locuss,
  title={LoCuSS: weak-lensing mass calibration of galaxy clusters},
  author={Okabe, Nobuhiro and Smith, Graham P},
  journal={MNRAS},
  volume={461},
  number={4},
  pages={3794--3821},
  year={2016},
  publisher={Oxford University Press}
}

@article{applegate2014weighing,
  title={Weighing the Giants--III. Methods and measurements of accurate galaxy cluster weak-lensing masses},
  author={Applegate, Douglas E and von der Linden, Anja and Kelly, Patrick L and Allen, Mark T and Allen, Steven W and Burchat, Patricia R and Burke, David L and Ebeling, Harald and Mantz, Adam and Morris, R Glenn},
  journal={MNRAS},
  volume={439},
  number={1},
  pages={48--72},
  year={2014},
  publisher={Oxford University Press}
}

@article{herbonnet2020cccp,
  title={CCCP and MENeaCS:(updated) weak-lensing masses for 100 galaxy clusters},
  author={Herbonnet, Ricardo and Sif{\'o}n, Crist{\'o}bal and Hoekstra, Henk and Bah{\'e}, Yannick and van Der Burg, Remco FJ and Melin, Jean-Baptiste and von Der Linden, Anja and Sand, David and Kay, Scott and Barnes, David},
  journal={MNRAS},
  volume={497},
  number={4},
  pages={4684--4703},
  year={2020},
  publisher={Oxford University Press}
}

@article{umetsu2016clash,
  title={CLASH: Joint analysis of strong-lensing, weak-lensing shear, and magnification data for 20 galaxy clusters},
  author={Umetsu, Keiichi and Zitrin, Adi and Gruen, Daniel and Merten, Julian and Donahue, Megan and Postman, Marc},
  journal={ApJ},
  volume={821},
  number={2},
  pages={116},
  year={2016},
  publisher={IOP Publishing}
}

@ARTICLE{2019A&A...621A..39Ettori,
       author = {{Ettori}, S. and {Ghirardini}, V. and {Eckert}, D. and {Pointecouteau}, E. and {Gastaldello}, F. and {Sereno}, M. and {Gaspari}, M. and {Ghizzardi}, S. and {Roncarelli}, M. and {Rossetti}, M.},
        title = "{Hydrostatic mass profiles in X-COP galaxy clusters}",
      journal = {\aap},
         year = 2019,
        month = jan,
       volume = {621},
          eid = {A39},
        pages = {A39},
          doi = {10.1051/0004-6361/201833323},
archivePrefix = {arXiv},
       eprint = {1805.00035},
 primaryClass = {astro-ph.CO},
       adsurl = {https://ui.adsabs.harvard.edu/abs/2019A&A...621A..39E}
}

@misc{lahav2022cosmologicalparameters2021,
      title={The Cosmological Parameters (2021)}, 
      author={Ofer Lahav and Andrew R Liddle},
      year={2022},
      eprint={2201.08666},
      archivePrefix={arXiv},
      primaryClass={astro-ph.CO},
      url={https://arxiv.org/abs/2201.08666}, 
}

@ARTICLE{2005A&A...435....1Pointecouteau,
   author = {{Pointecouteau}, E. and {Arnaud}, M. and {Pratt}, G.~W.},
    title = "{The structural and scaling properties of nearby galaxy clusters. I. The universal mass profile}",
  journal = {\aap},
     year = 2005,
   volume = 435,
    pages = {1},
      doi = {10.1051/0004-6361:20042569},
   adsurl = {https://ui.adsabs.harvard.edu/abs/2005A&A...435....1P}
}

@ARTICLE{2009A&A...498..361Pratt,
   author = {{Pratt}, G.~W. and {Croston}, J.~H. and {Arnaud}, M. and {B{\"o}hringer}, H.},
    title = "{Galaxy cluster X-ray luminosity scaling relations from a representative local sample (REXCESS)}",
  journal = {\aap},
     year = 2009,
   volume = 498,
    pages = {361},
      doi = {10.1051/0004-6361/200810994},
   adsurl = {https://ui.adsabs.harvard.edu/abs/2009A&A...498..361P}
}

@ARTICLE{2009ApJ...692.1060Vikhlinin,
   author = {{Vikhlinin}, A. and {Kravtsov}, A.~V. and {Burenin}, R.~A. and {Ebeling}, H.
             and {Forman}, W.~R. and {Hornstrup}, A. and {Jones}, C. and {Murray}, S.~S.
             and {Nagai}, D. and {Quintana}, H. and {Voevodkin}, A.},
    title = "{Chandra Cluster Cosmology Project III: Cosmological Parameter Constraints}",
  journal = {\apj},
     year = 2009,
   volume = 692,
    pages = {1060},
      doi = {10.1088/0004-637X/692/2/1060},
   adsurl = {https://ui.adsabs.harvard.edu/abs/2009ApJ...692.1060V}
}

@ARTICLE{2024A&A...682A.157Lebeau,
       author = {{Lebeau}, Th{\'e}o and {Sorce}, Jenny G. and {Aghanim}, Nabila and {Hern{\'a}ndez-Mart{\'\i}nez}, Elena and {Dolag}, Klaus},
        title = "{Mass bias in clusters of galaxies: Projection effects on the case study of Virgo replica}",
      journal = {\aap},
         year = 2024,
        month = feb,
       volume = {682},
          eid = {A157},
        pages = {A157},
          doi = {10.1051/0004-6361/202347511},
archivePrefix = {arXiv},
       eprint = {2310.02326},
 primaryClass = {astro-ph.CO},
       adsurl = {https://ui.adsabs.harvard.edu/abs/2024A&A...682A.157L}
}

@ARTICLE{2024A&A...689A..19Lebeau,
       author = {{Lebeau}, Th{\'e}o and {Ettori}, Stefano and {Aghanim}, Nabila and {Sorce}, Jenny G.},
        title = "{Can the splashback radius be an observable boundary of galaxy clusters?}",
      journal = {\aap},
         year = 2024,
        month = sep,
       volume = {689},
          eid = {A19},
        pages = {A19},
          doi = {10.1051/0004-6361/202450146},
archivePrefix = {arXiv},
       eprint = {2403.18648},
 primaryClass = {astro-ph.CO},
       adsurl = {https://ui.adsabs.harvard.edu/abs/2024A&A...689A..19L}
}

@ARTICLE{2025A&A...694A.182Adam,
       author = {{Adam}, R. and {Eynard-Machet}, T. and {Bartalucci}, I. and {Cherouvrier}, D. and {Clerc}, N. and {Di Mascolo}, L. and {Dupourqu{\'e}}, S. and {Ferrari}, C. and {Mac{\'\i}as-P{\'e}rez}, J.-F. and {Pointecouteau}, E. and {Pratt}, G.~W.},
        title = "{PITSZI: Probing intra-cluster medium turbulence with Sunyaev{\textendash}Zel'dovich imaging: Application to the triple merging cluster MACS J0717.5+3745}",
      journal = {\aap},
         year = 2025,
        month = feb,
       volume = {694},
          eid = {A182},
        pages = {A182},
          doi = {10.1051/0004-6361/202452342},
archivePrefix = {arXiv},
       eprint = {2409.14804},
 primaryClass = {astro-ph.CO},
       adsurl = {https://ui.adsabs.harvard.edu/abs/2025A&A...694A.182A}
}

@ARTICLE{2025ApJ...993L..11XRISM_comp,
       author = {{XRISM Collaboration} and {Audard}, Marc and {Awaki}, Hisamitsu and {Ballhausen}, Ralf and {Bamba}, Aya and {Behar}, Ehud and {Boissay-Malaquin}, Rozenn and {Brenneman}, Laura and {Brown}, Gregory V. and {Corrales}, Lia and {Costantini}, Elisa and {Cumbee}, Renata and {Diaz Trigo}, Maria and {Done}, Chris and {Dotani}, Tadayasu and {Ebisawa}, Ken and {Eckart}, Megan E. and {Eckert}, Dominique and {Eguchi}, Satoshi and {Enoto}, Teruaki and {Ezoe}, Yuichiro and {Foster}, Adam and {Fujimoto}, Ryuichi and {Fujita}, Yutaka and {Fukazawa}, Yasushi and {Fukushima}, Kotaro and {Furuzawa}, Akihiro and {Gallo}, Luigi and {Garc{\'\i}a}, Javier A. and {Gu}, Liyi and {Guainazzi}, Matteo and {Hagino}, Kouichi and {Hamaguchi}, Kenji and {Hatsukade}, Isamu and {Hayashi}, Katsuhiro and {Hayashi}, Takayuki and {Hell}, Natalie and {Hodges-Kluck}, Edmund and {Hornschemeier}, Ann and {Ichinohe}, Yuto and {Ishi}, Daiki and {Ishida}, Manabu and {Ishikawa}, Kumi and {Ishisaki}, Yoshitaka and {Kaastra}, Jelle and {Kallman}, Timothy and {Kara}, Erin and {Katsuda}, Satoru and {Kanemaru}, Yoshiaki and {Kelley}, Richard and {Kilbourne}, Caroline and {Kitamoto}, Shunji and {Kobayashi}, Shogo and {Kohmura}, Takayoshi and {Kubota}, Aya and {Leutenegger}, Maurice and {Loewenstein}, Michael and {Maeda}, Yoshitomo and {Markevitch}, Maxim and {Matsumoto}, Hironori and {Matsushita}, Kyoko and {McCammon}, Dan and {McNamara}, Brian and {Mernier}, Fran{\c{c}}ois and {Miller}, Eric D. and {Miller}, Jon M. and {Mitsuishi}, Ikuyuki and {Mizumoto}, Misaki and {Mizuno}, Tsunefumi and {Mori}, Koji and {Mukai}, Koji and {Murakami}, Hiroshi and {Mushotzky}, Richard and {Nakajima}, Hiroshi and {Nakazawa}, Kazuhiro and {Ness}, Jan-Uwe and {Nobukawa}, Kumiko and {Nobukawa}, Masayoshi and {Noda}, Hirofumi and {Odaka}, Hirokazu and {Ogawa}, Shoji and {Ogorza{\l}ek}, Anna and {Okajima}, Takashi and {Ota}, Naomi and {Paltani}, Stephane and {Petre}, Robert and {Plucinsky}, Paul and {Porter}, Frederick S. and {Pottschmidt}, Katja and {Sato}, Kosuke and {Sato}, Toshiki and {Sawada}, Makoto and {Seta}, Hiromi and {Shidatsu}, Megumi and {Simionescu}, Aurora and {Smith}, Randall and {Suzuki}, Hiromasa and {Szymkowiak}, Andrew and {Takahashi}, Hiromitsu and {Takeo}, Mai and {Tamagawa}, Toru and {Tamura}, Keisuke and {Tanaka}, Takaaki and {Tanimoto}, Atsushi and {Tashiro}, Makoto and {Terada}, Yukikatsu and {Terashima}, Yuichi and {Tsuboi}, Yohko and {Tsujimoto}, Masahiro and {Tsunemi}, Hiroshi and {Tsuru}, Takeshi and {T{\"u}mer}, Ay{\textcommabelow s}eg{\"u}l and {Uchida}, Hiroyuki and {Uchida}, Nagomi and {Uchida}, Yuusuke and {Uchiyama}, Hideki and {Ueda}, Shutaro and {Ueda}, Yoshihiro and {Uno}, Shinichiro and {Vink}, Jacco and {Watanabe}, Shin and {Williams}, Brian J. and {Yamada}, Satoshi and {Yamada}, Shinya and {Yamaguchi}, Hiroya and {Yamaoka}, Kazutaka and {Yamasaki}, Noriko and {Yamauchi}, Makoto and {Yamauchi}, Shigeo and {Yaqoob}, Tahir and {Yoneyama}, Tomokage and {Yoshida}, Tessei and {Yukita}, Mihoko and {Zhuravleva}, Irina and {Cui}, Weiguang and {Ettori}, Stefano and {Grayson}, Skylar and {Heinrich}, Annie and {McCall}, Hannah and {Nelson}, Dylan and {Okabe}, Nobuhiro and {Omiya}, Yuki and {Sarkar}, Arnab and {Scannapieco}, Evan and {Sun}, Ming and {Tanaka}, Keita and {Truong}, Nhut and {Wik}, Daniel R. and {Zhang}, Congyao and {Zuhone}, John},
        title = "{Comparing XRISM Cluster Velocity Dispersions with Predictions from Cosmological Simulations: Are Feedback Models Too Ejective?}",
      journal = {\apjl},
         year = 2025,
        month = nov,
       volume = {993},
       number = {1},
          eid = {L11},
        pages = {L11},
          doi = {10.3847/2041-8213/ae100c},
archivePrefix = {arXiv},
       eprint = {2510.06322},
 primaryClass = {astro-ph.GA},
       adsurl = {https://ui.adsabs.harvard.edu/abs/2025ApJ...993L..11X}
}

@ARTICLE{2019A&A...629A.143Clerc,
       author = {{Clerc}, Nicolas and {Cucchetti}, Edoardo and {Pointecouteau}, Etienne and {Peille}, Philippe},
        title = "{Towards mapping turbulence in the intra-cluster medium. I. Sample variance in spatially-resolved X-ray line diagnostics}",
      journal = {\aap},
         year = 2019,
        month = sep,
       volume = {629},
          eid = {A143},
        pages = {A143},
          doi = {10.1051/0004-6361/201935676},
archivePrefix = {arXiv},
       eprint = {1904.06248},
 primaryClass = {astro-ph.CO},
       adsurl = {https://ui.adsabs.harvard.edu/abs/2019A&A...629A.143C}
}

@ARTICLE{2024A&A...686A..41Beaumont,
       author = {{Beaumont}, S. and {Molin}, A. and {Clerc}, N. and {Pointecouteau}, E. and {Vanel}, M. and {Cucchetti}, E. and {Peille}, P. and {Pajot}, F.},
        title = "{Toward mapping turbulence in the intra-cluster medium. III. Constraints on the turbulent power spectrum with Athena/X-IFU}",
      journal = {\aap},
         year = 2024,
        month = jun,
       volume = {686},
          eid = {A41},
        pages = {A41},
          doi = {10.1051/0004-6361/202348937},
archivePrefix = {arXiv},
       eprint = {2403.08601},
 primaryClass = {astro-ph.CO},
       adsurl = {https://ui.adsabs.harvard.edu/abs/2024A&A...686A..41B}
}

@ARTICLE{2025A&A...702A.215Molin,
       author = {{Molin}, A. and {Dupourqu{\'e}}, S. and {Clerc}, N. and {Pointecouteau}, E. and {Pajot}, F. and {Cucchetti}, E.},
        title = "{Toward mapping turbulence in the intracluster medium: IV. Using NewAthena/X-IFU and simulation-based inference to constrain turbulence}",
      journal = {\aap},
         year = 2025,
        month = oct,
       volume = {702},
          eid = {A215},
        pages = {A215},
          doi = {10.1051/0004-6361/202555585},
archivePrefix = {arXiv},
       eprint = {2505.14378},
 primaryClass = {astro-ph.CO},
       adsurl = {https://ui.adsabs.harvard.edu/abs/2025A&A...702A.215M}
}

@article{sorce2026ii,
  title={II-A hydrodynamical CLONE of the Virgo cluster to confront observed and synthetic galaxy population twins in a dense environment},
  author={Sorce, Jenny G and McGee, Sean L and Dubois, Yohan and Blaizot, J{\'e}r{\'e}my and Knebe, Alexander and Yepes, Gustavo},
  journal={arXiv preprint arXiv:2603.23606},
  year={2026}
}

@ARTICLE{2025arXiv250902068Aymerich,
       author = {{Aymerich}, G. and {Grandis}, S. and {Douspis}, M. and {Pratt}, G.~W. and {Salvati}, L. and {Andrade-Santos}, F. and {Bocquet}, S. and {Costanzi}, M. and {Forman}, W.~R. and {Jones}, C. and {Aguena}, M. and {Andrade-Oliveira}, F. and {Bacon}, D. and {Brooks}, D. and {Burke}, D.~L. and {Carretero}, J. and {da Costa}, L.~N. and {da Silva Pereira}, M.~E. and {Davis}, T.~M. and {De Vicente}, J. and {Desai}, S. and {Diehl}, H.~T. and {Doel}, P. and {Everett}, S. and {Flaugher}, B. and {Frieman}, J. and {Gaztanaga}, E. and {Gruen}, D. and {Gutierrez}, G. and {Hinton}, S.~R. and {Hollowood}, D.~L. and {Honscheid}, K. and {James}, D.~J. and {Lee}, S. and {Marshall}, J.~L. and {Mena-Fern{\'a}ndez}, J. and {Miquel}, R. and {Mohr}, J.~J. and {Ogando}, R.~L.~C. and {Plazas Malag{\'o}n}, A.~A. and {Porredon}, A. and {Prat}, J. and {Romer}, A.~K. and {Samuroff}, S. and {Sanchez}, E. and {Sanchez Cid}, D. and {Smith}, M. and {Suchyta}, E. and {Swanson}, M.~E.~C. and {Tucker}, D.~L. and {Weaverdyck}, N. and {Weller}, J. and {Yamamoto}, M.},
        title = "{Cosmological constraints from the Planck cluster catalogue with DES shear profiles and Chandra observations}",
      journal = {arXiv e-prints},
         year = 2025,
        month = sep,
          eid = {arXiv:2509.02068},
        pages = {arXiv:2509.02068},
          doi = {10.48550/arXiv.2509.02068},
archivePrefix = {arXiv},
       eprint = {2509.02068},
 primaryClass = {astro-ph.CO},
       adsurl = {https://ui.adsabs.harvard.edu/abs/2025arXiv250902068A}
}

@article{2020MNRAS.495..864Angelinelli,
    author = {Angelinelli, M. and Vazza, F. and Giocoli, C. and Ettori, S. and Jones, T.~W. and Brunetti, G. and Br{\"u}ggen, M. and Eckert, D.},
    title = {Turbulent pressure support and hydrostatic mass bias in the intracluster medium},
    journal = {MNRAS},
    volume = {495},
    pages = {864--885},
    year = {2020},
    doi = {10.1093/mnras/staa975}
}

@article{2014MNRAS.442..521Shi,
    author = {Shi, Xun and Komatsu, Eiichiro},
    title = {Analytical model for non-thermal pressure in galaxy clusters},
    journal = {MNRAS},
    volume = {442},
    pages = {521--532},
    year = {2014},
    doi = {10.1093/mnras/stu858}
}

@article{2016MNRAS.455.2936Shi,
    author = {Shi, Xun and Komatsu, Eiichiro and Nagai, Daisuke and Lau, Erwin T.},
    title = {Analytical model for non-thermal pressure in galaxy clusters -- {III}. {R}emoving the hydrostatic mass bias},
    journal = {MNRAS},
    volume = {455},
    pages = {2936--2944},
    year = {2016},
    doi = {10.1093/mnras/stv2504}
}

@ARTICLE{2025NatAs...9...36Cruise,
       author = {{Cruise}, Mike and {Guainazzi}, Matteo and {Aird}, James and {Carrera}, Francisco J. and {Costantini}, Elisa and {Corrales}, Lia and {Dauser}, Thomas and {Eckert}, Dominique and {Gastaldello}, Fabio and {Matsumoto}, Hironori and {Osten}, Rachel and {Petrucci}, Pierre-Olivier and {Porquet}, Delphine and {Pratt}, Gabriel W. and {Rea}, Nanda and {Reiprich}, Thomas H. and {Simionescu}, Aurora and {Spiga}, Daniele and {Troja}, Eleonora},
        title = "{The NewAthena mission concept in the context of the next decade of X-ray astronomy}",
      journal = {Nature Astronomy},
         year = 2025,
        month = jan,
       volume = {9},
        pages = {36-44},
          doi = {10.1038/s41550-024-02416-3},
archivePrefix = {arXiv},
       eprint = {2501.03100},
 primaryClass = {astro-ph.IM},
       adsurl = {https://ui.adsabs.harvard.edu/abs/2025NatAs...9...36C}
}

@ARTICLE{2022A&A...657L...1Ettori,
       author = {{Ettori}, S. and {Eckert}, D.},
        title = "{Tracing the non-thermal pressure and hydrostatic bias in galaxy clusters}",
      journal = {\aap},
         year = 2022,
        month = jan,
       volume = {657},
          eid = {L1},
        pages = {L1},
          doi = {10.1051/0004-6361/202142638},
archivePrefix = {arXiv},
       eprint = {2112.07554},
 primaryClass = {astro-ph.CO},
       adsurl = {https://ui.adsabs.harvard.edu/abs/2022A&A...657L...1E}
}

@ARTICLE{2025A&A...697A..17Bartalesi,
       author = {{Bartalesi}, T. and {Ettori}, S. and {Nipoti}, C.},
        title = "{Searching for rotation in X-COP galaxy clusters}",
      journal = {\aap},
         year = 2025,
        month = may,
       volume = {697},
          eid = {A17},
        pages = {A17},
          doi = {10.1051/0004-6361/202553720},
archivePrefix = {arXiv},
       eprint = {2503.13612},
 primaryClass = {astro-ph.CO},
       adsurl = {https://ui.adsabs.harvard.edu/abs/2025A&A...697A..17B}
}

\appendix

\twocolumn 

\section{Regions studied in the 3D case}

\begin{figure}[h!]
    \centering
    \includegraphics[trim= 400 20 600 20,clip,width=0.9\linewidth]{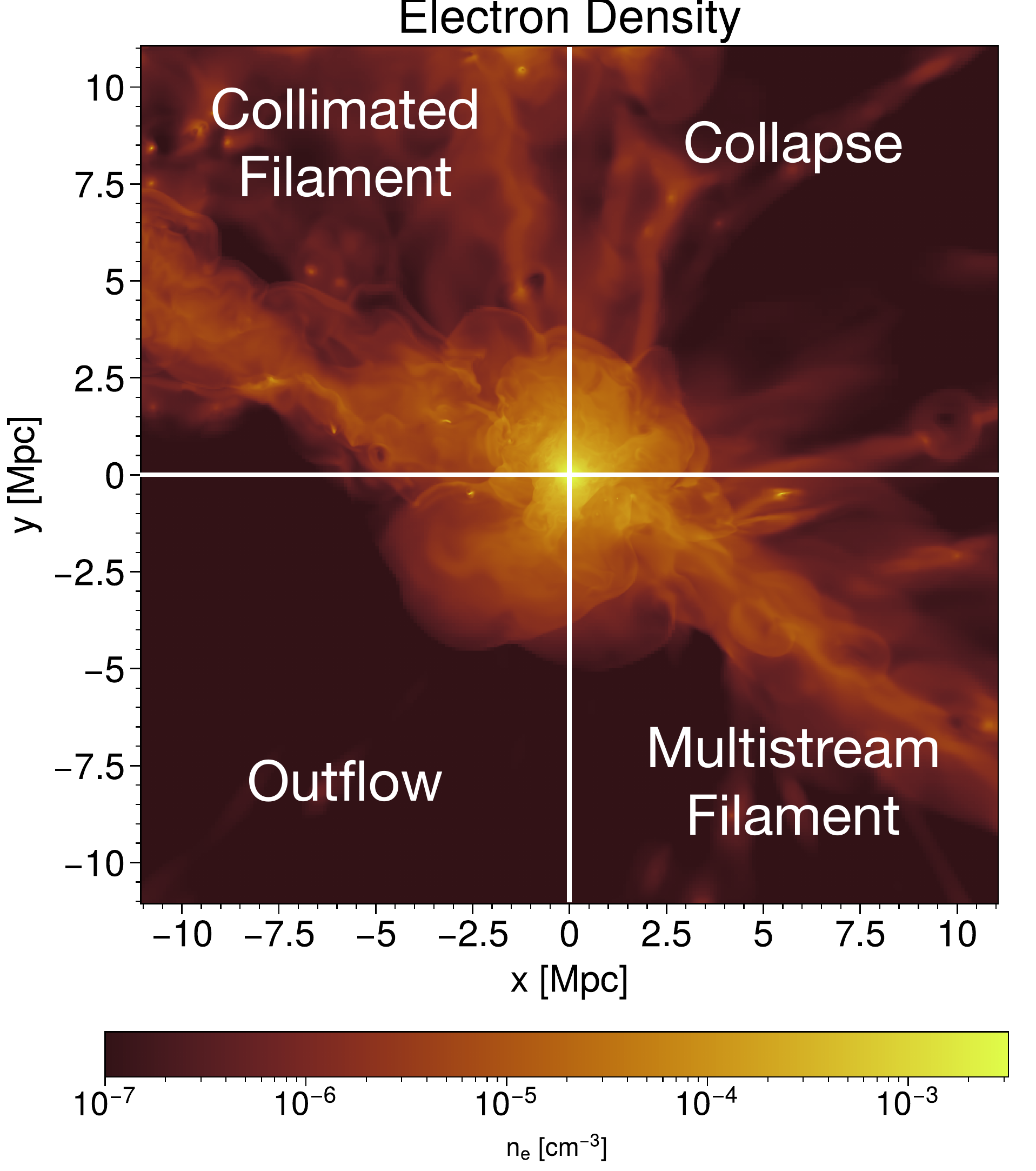}
    \caption{Slice of the electron density taken along the z axis of the simulation showing the regions studied in Sect.~\ref{sec4: 3d mass}. This figure is extracted from Fig. 1 of \citet{2025A&A...704A..14Lebeau}.}
    \label{app:sectors_ne}
\end{figure}

\section{3D pressure and electron density radial profiles}

\begin{figure}[h!]
    \centering
    \includegraphics[trim= 0 20 0 70,clip,width=0.85\linewidth]{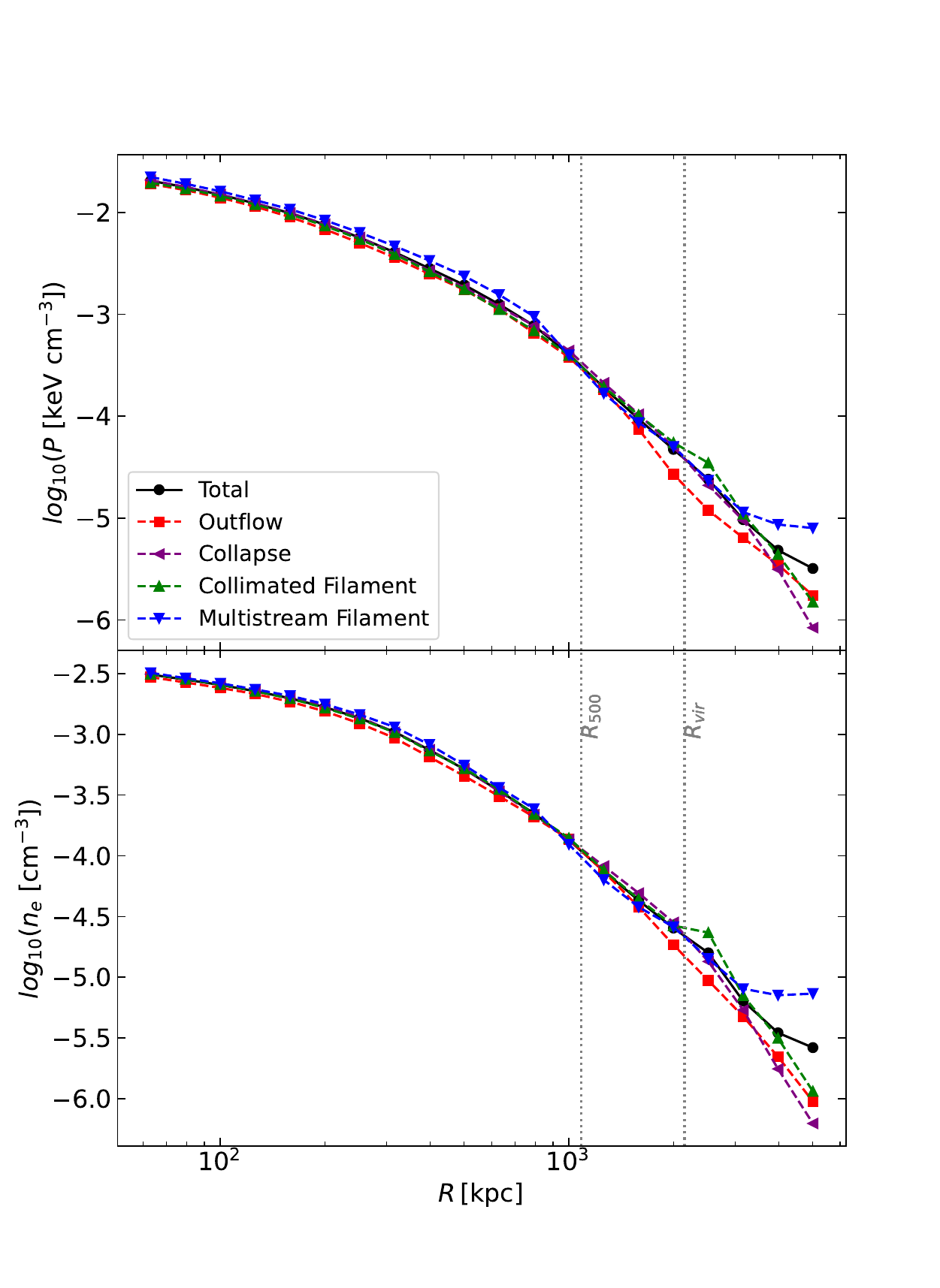}
    \caption{3D pressure (top) and electron density (bottom) radial profiles estimated in the regions studied in Sect.~\ref{sec4: 3d mass}.}
    \label{app:p_ne_3D_profs}
\end{figure}

\section{Mass bias comparisons}
\label{app:sec:mass comp}

\begin{figure}[H]
    \centering
    \includegraphics[trim= 0 0 0 20,clip,width=1\linewidth]{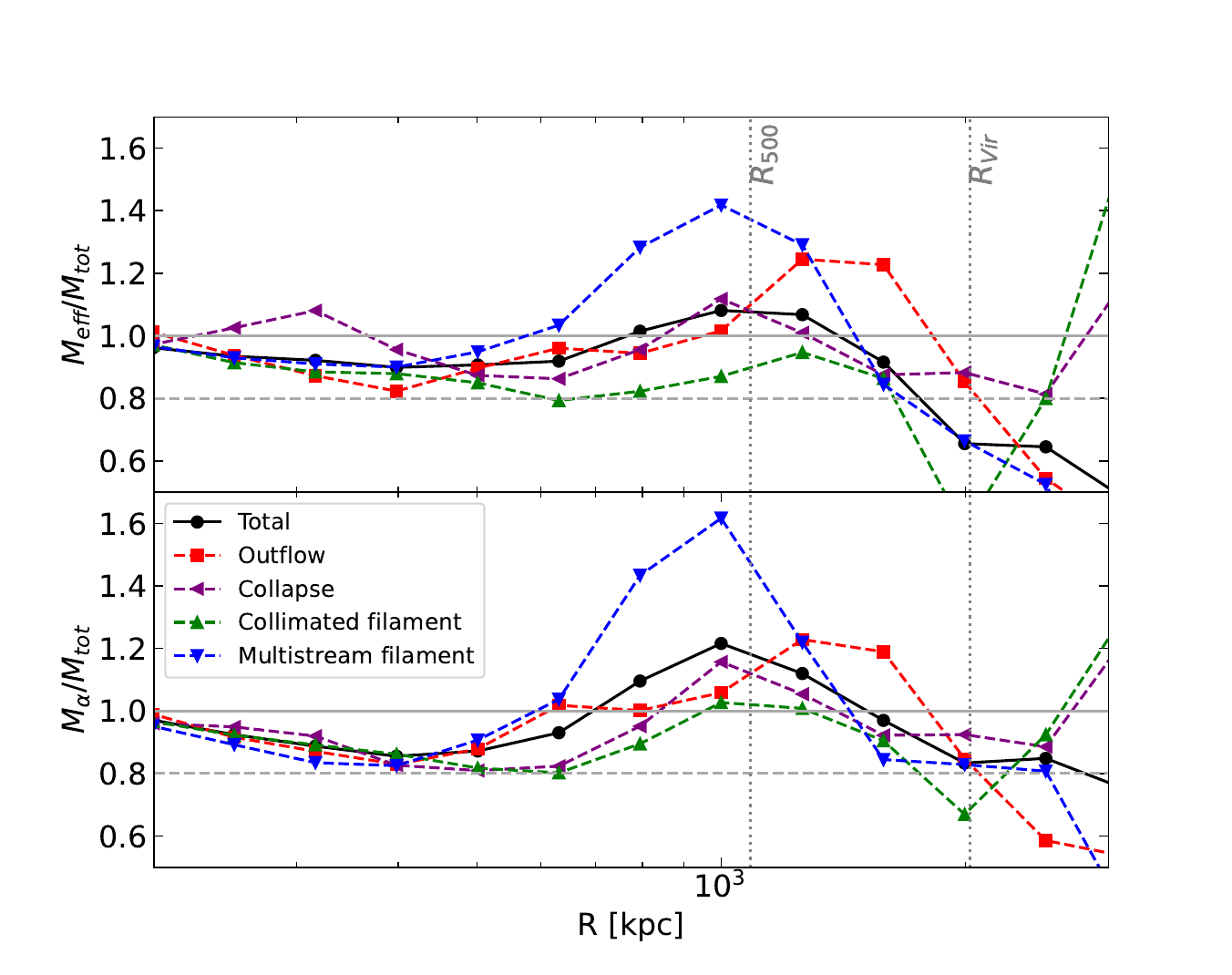}
    \caption{Ratios of $M_{\rm eff}$ (top) and $M_{\rm \alpha}$ (bottom) over $M_{\rm HE}$.}
    \label{app:M_ratio_comp_3D}
\end{figure}

\section{Estimating $v_{\rm t}$ from projections}
\label{app:v_t}

\begin{figure}[h!]
    \centering
    \includegraphics[trim= 300 0 450 0,clip,width=0.9\linewidth]{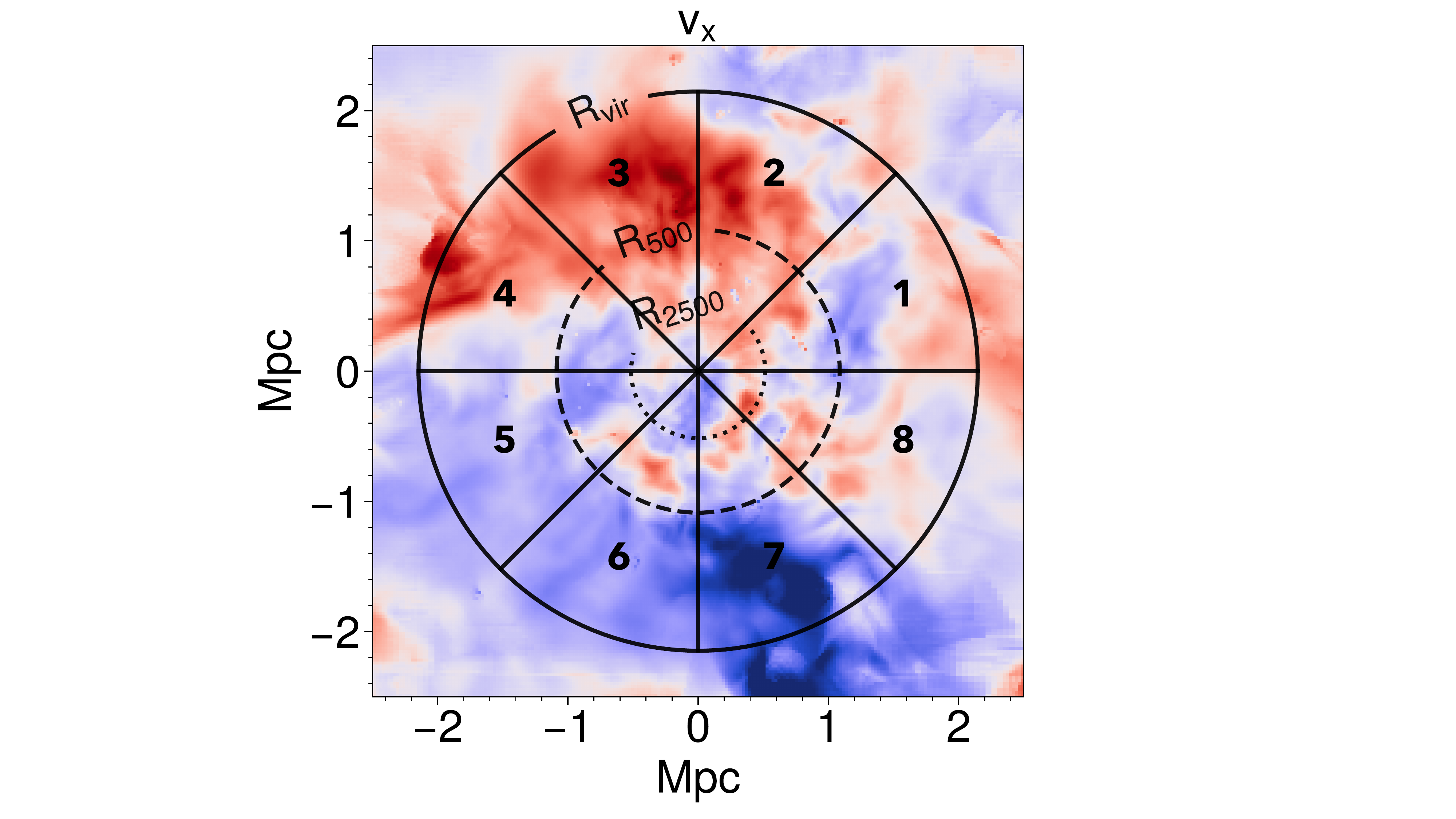}
    \caption{Quadrants used to estimate $v_{\rm t}$ on the $v_{\rm x,ew}$ projection as an example.}
    \label{app:vt_quadrants}
\end{figure}

The tangential velocity, $v_t$, cannot be measured directly. To overcome this, we make the assumption, following \cite{2018PASJ...70...51Ota}, that if the cluster has a global bulk rotation, it induces a slightly higher sightline velocity in a region of the cluster and a smaller one in the opposite region. We thus divide the sightline velocity maps into eight sectors, as shown in Fig.~\ref{app:vt_quadrants}, numbered from 1 to 8 in trigonometric order, each with an aperture of $\pi/4$ in the [0,$2\pi$] range. We thus have four pairs of opposite sectors: 1-5, 2-6, 3-7 and 4-8. Therefore, the maximum absolute velocity difference among the pairs, $\max(\Delta v_\mathrm{LOS})$, is assimilated to $v_t$.

\onecolumn

\section{Table of the masses}
\label{app:masses section}

\begin{table*}[h!]
    \large
    \centering
    \caption{Mass estimated from the 3D radial profiles (entire cluster) and from the projected quantities along the four sightlines, using both emission-weighted (ew) and mass-weighted (mw) projections. All masses are in units of $10^{14}~M_{\odot}$.}
    \begin{tabular}{ c c c c c c c c c c }
    \hline \hline \hline
    \multirow{3}{*}{Mass terms} & \multirow{1}{*}{3D} & \multicolumn{8}{c}{2D}\\
    & & \multicolumn{4}{c}{Emission-weighted} & \multicolumn{4}{c}{Mass-weighted} \\
    & & $x$ & $y$ & $z$ & $cen$ & $x$ & $y$ & $z$ & $cen$ \\
    \hline \hline 
    & \multicolumn{9}{c}{Masses within $R_{500}$=1087~kpc (in $10^{14}M_{\odot}$)} \\
     \noalign{\vskip 0.1cm}    
    $M_{\rm HE}$ & 3.39 & 10.4 & 8.92 & 6.65 & 4.55 & 10.4 & 8.92 & 6.65 & 4.55 \\
    $M_{\alpha}$ & 4.03 & 11.0 & 9.40 & 7.10 & 4.92 & 11.8 & 9.74 & 7.55 & 4.58 \\
    $M_{\rm disp}$ & $-$0.24 & 0.26 & 0.21 & 0.21 & 0.15 & 0.61 & 0.38 & 0.45 & 0.01 \\
    $M_{\rm rot}$ & 0.24 & 0.02 & 0.0002 & 0.06 & 0.05 & 0.006 & 0.004 & 0.01 & 0.007 \\
    $M_{\rm eff}$ & 3.39 & 10.7 & 9.13 & 6.92 & 4.76 & 11.0 & 9.30 & 7.11 & 4.57 \\
    $M_{\rm tot}$ & 3.56\\
    \hline \hline 
    & \multicolumn{9}{c}{Masses within $R_{\rm vir}$=2147~kpc (in $10^{14}M_{\odot}$)} \\
     \noalign{\vskip 0.1cm}    
    $M_{\rm HE}$ & 4.06 & 6.80 & 6.55 & 4.65 & 4.29 & 6.80 & 6.55 & 4.65 & 4.29 \\
    $M_{\alpha}$ & 5.78 & 8.03 & 7.40 & 5.62 & 4.83 & 8.80 & 8.06 & 6.20 & 4.45 \\
    $M_{\rm disp}$ & $-$0.13 & 0.60 & 0.46 & 0.53 & 0.26 & 1.12 & 0.88 & 0.94 & 0.07 \\
    $M_{\rm rot}$ & 0.46 & 1.46 & 0.20 & 0.25 & 0.12 & 1.12 & 0.27 & 0.20 & 0.09 \\
    $M_{\rm eff}$ & 4.39 & 8.87 & 7.21 & 5.43 & 4.67 & 9.04 & 7.70 & 5.80 & 4.45 \\
    $M_{\rm tot}$ & 6.58\\
    \hline
    \hline
    \hline
    
    \end{tabular}
    \label{app: masses tab}
\end{table*}

\section{Table of the mass biases}
\label{app:bias section}

\begin{table*}[h!]
    \large
    \centering
    \renewcommand{\arraystretch}{1.3}
    \caption{Mass biases $M/M_{\rm tot}$ estimated from the 3D radial profiles (entire cluster) and from the projected quantities along the four sightlines, using both emission-weighted (ew) and mass-weighted (mw) projections. The values are computed from the masses in Tab.~\ref{app: masses tab}.}
    \begin{tabular}{ c c c c c c c c c c }
    \hline \hline \hline
    \multirow{3}{*}{Mass bias} & \multirow{1}{*}{3D} & \multicolumn{8}{c}{2D}\\
    & & \multicolumn{4}{c}{Emission-weighted} & \multicolumn{4}{c}{Mass-weighted} \\
    & & $x$ & $y$ & $z$ & $cen$ & $x$ & $y$ & $z$ & $cen$ \\
    \hline \hline 
    & \multicolumn{9}{c}{Mass biases within $R_{500}$=1087~kpc} \\
     \noalign{\vskip 0.1cm}    
    $M_{\rm HE}/M_{\rm tot}$ & 0.95 & 2.92 & 2.51 & 1.87 & 1.28 & 2.92 & 2.51 & 1.87 & 1.28 \\
    $M_{\rm \alpha}/M_{\rm tot}$ & 1.13 & 3.09 & 2.64 & 1.99 & 1.38 & 3.31 & 2.74 & 2.12 & 1.29 \\
    $M_{\rm eff}/M_{\rm tot}$ & 0.95 & 3.01 & 2.56 & 1.94 & 1.34 & 3.09 & 2.61 & 2.00 & 1.28 \\
    \hline \hline 
    & \multicolumn{9}{c}{Mass biases within $R_{vir}$=2147~kpc} \\
     \noalign{\vskip 0.1cm}    
    $M_{\rm HE}/M_{\rm tot}$ & 0.62 & 1.03 & 1.00 & 0.71 & 0.65 & 1.03 & 1.00 & 0.71 & 0.65 \\
    $M_{\rm \alpha}/M_{\rm tot}$ & 0.88 & 1.22 & 1.12 & 0.85 & 0.73 & 1.34 & 1.22 & 0.94 & 0.68 \\
    $M_{\rm eff}/M_{\rm tot}$ & 0.67 & 1.35 & 1.10 & 0.83 & 0.71 & 1.37 & 1.17 & 0.88 & 0.68 \\
    \hline
    \hline
    \hline
    
    \end{tabular}
    \label{app: bias tab}
\end{table*}

\end{document}